\documentclass[fleqn,usenatbib]{mnras}

\usepackage{newtxtext,newtxmath}
\usepackage[T1]{fontenc}
\usepackage{ae,aecompl}

\usepackage{graphicx}	
\usepackage{amssymb}	

\usepackage{hyperref}
\usepackage{xspace}
\usepackage{url}
\usepackage{longtable}
\usepackage[flushleft]{threeparttablex}
\usepackage{natbib,twoopt}

\def\simeq{
\mathrel{\raise.3ex\hbox{$\sim$}\mkern-14mu\lower0.4ex\hbox{$-$}}
}

\def\ltsima{$\; \buildrel < \over \sim \;$}
\def\simlt{\lower.5ex\hbox{\ltsima}}
\def\gtsima{$\; \buildrel > \over \sim \;$}
\def\simgt{\lower.5ex\hbox{\gtsima}}

\def\msun{{\rm M_{\odot}}}

\def\be{\begin{equation}}
\def\ee{\end{equation}}

\def\del#1{{}}
\def\ltsima{$\; \buildrel < \over \sim \;$}
\def\simlt{\lower.5ex\hbox{\ltsima}}
\def\gtsima{$\; \buildrel > \over \sim \;$}
\def\simgt{\lower.5ex\hbox{\gtsima}}

\title[Intermittent driving of AGN outflows]{Intermittent AGN episodes drive outflows with a large spread of observable loading factors}

\author[K. Zubovas, E. Nardini]{Kastytis Zubovas$^{1,2,\star}$ and Emanuele Nardini$^{3,4}$ \\
  $^{1}$Center for Physical Sciences and Technology, Saul\.{e}tekio al. 3, Vilnius LT-10257, Lithuania\\
  $^{2}$Vilnius University Observatory, Saul\.{e}tekio al. 3, Vilnius LT-10257, Lithuania\\
  $^{3}$Dipartimento di Fisica e Astronomia, Universit\`a di Firenze, via G. Sansone 1, I-50019 Sesto Fiorentino, Firenze, Italy \\
  $^{4}$INAF - Osservatorio Astrofisico di Arcetri, Largo Enrico Fermi 5, I-50125 Firenze, Italy \\
  $^{\star}$ {E-mail:~} {\rm kastytis.zubovas@ftmc.lt} }

\date{Accepted XXX. Received YYY; in original form ZZZ}

\pubyear{2019}

\begin{document}
\label{firstpage}
\pagerange{\pageref{firstpage}--\pageref{lastpage}}
\maketitle

\begin{abstract}

The properties of large-scale galactic outflows, such as their kinetic energy and momentum rates, correlate with the luminosity of the active galactic nucleus (AGN). This is well explained by the wind-driven outflow model, where a fraction of the AGN luminosity drives the outflow. However, significant departures from these correlations have been observed in a number of galaxies. This may happen because AGN luminosity varies on a much shorter timescale ($\sim 10^4-10^5$~yr) than outflow properties do ($\sim 10^6$~yr). We investigate the effect of AGN luminosity variations on outflow properties using 1D numerical simulations. This effect can explain the very weak outflow in PDS 456: if its nucleus is currently much brighter than the long-term average luminosity, the outflow has not had time to react to this luminosity change. Conversely, the outflow in Mrk 231 is consistent with being driven by an almost continuous AGN, while IRAS F11119+3257 represents an intermediate case between the two. Considering a population of AGN, we find that very low momentum loading factors $\dot{p}_{\rm out} < L_{\rm AGN}/c$ should be seen in a significant fraction of objects - up to $15\%$ depending on the properties of AGN variability and galaxy gas fraction. The predicted distribution of loading factors is consistent with the available observational data. We discuss how this model might help constrain the duty cycles of AGN during the period of outflow inflation, implications for multiphase and spatially distinct outflows, and suggest ways of improving AGN prescriptions in numerical simulations.
\end{abstract}




\section{Introduction} \label{sec:intro}

Supermassive black holes (SMBHs) are known to exist in the majority of, if not all, galaxies \citep{Heckman2014ARAA}. For a small fraction of its lifetime, each SMBH accretes material rapidly and becomes visible as an active galactic nucleus (AGN). Many AGN show evidence of fast, quasi-relativistic winds with velocities $v_{\rm w} \sim 0.1c$ that carry a significant fraction $\eta \sim 0.05$ of the AGN luminosity as kinetic power \citep{Pounds2003MNRASa, Tombesi2014MNRAS}. These winds are believed to drive large-scale outflows that sweep gas out of galaxies, quenching their star formation \citep{Feruglio2010AA, Sturm2011ApJ, Tombesi2015Natur}. This mechanism is responsible for the high-mass cutoff in the galaxy mass function, as shown in many semi-analytical \citep[e.g.,][]{Bower2006MNRAS, Croton2006MNRAS} and hydrodynamical simulations that test the effects of AGN feedback \citep[e.g.,][]{Sijacki2007MNRAS, Puchwein2013MNRAS, Dubois2014MNRAS, Vogelsberger2014MNRAS, Schaye2015MNRAS, Tremmel2019MNRAS}.

Many outflow properties, such as the mass flow rate $\dot{M}_{\rm out}$, velocity $v_{\rm out}$, momentum rate $\dot{p}_{\rm out} = \dot{M}_{\rm out} v_{\rm out}$ and kinetic energy rate $\dot{E}_{\rm out} = \dot{M}_{\rm out} v_{\rm out}^2/2$ correlate with the luminosity of the driving AGN \citep{Cicone2014AA, Fiore2017AA, Gonzalez2017ApJ, Fluetsch2019MNRAS, Lutz2020AA}. These correlations have been predicted by the wind-driven outflow model \citep{King2010MNRASa, Zubovas2012ApJ, King2015ARAA}, which is based on the assumption that large-scale outflows are inflated when the SMBH mass reaches a critical value, approximately given by the $M-\sigma$ relation \citep[e.g.,][]{McConnell2013ApJ}; once this happens, the AGN wind transfers most of its energy to the interstellar medium (ISM), rather than losing it to cooling processes.

In recent years, with the increased availability of outflow data, it became clear that many outflows have mass, momentum and energy rates that fall below the analytically predicted correlations \citep[e.g.][]{Fiore2017AA,Harrison2018NatAs}. One interpretation of this phenomenon is that in these galaxies, a large part of the energy carried by the AGN wind does not couple to the ISM \citep{Smith2019ApJ}, perhaps due to being efficiently radiated away. In extreme cases, it has been suggested that some outflows might be momentum-conserving rather than energy-conserving \citep{Bischetti2019AA, Sirressi2019MNRAS, Reeves2019ApJ}. A momentum-conserving outflow has a momentum loading factor $\dot{p}_{\rm out}c/L_{\rm AGN} \sim 1$ and $\dot{E}_{\rm out}/L_{\rm AGN} < 10^{-3}$, consistent with these observations. A possible test of this interpretation is given by the prediction that a rapidly cooling AGN wind would emit a large amount of radiation, $L_{\rm cool} \simeq 0.05 L_{\rm AGN}$ \citep{Bourne2013MNRAS, Nims2015MNRAS}. This is generally not observed, although in some cases, evidence suggests the presence of cooling wind \citep{Pounds2013MNRAS}. Additionally, the lack of detailed information about the distribution of outflowing gas in different phases (molecular, neutral and ionized), as well as the averaging of mass flow rates over the outflow lifetime, might lead to much lower estimates of momentum flow rates than the simple analytical prediction \citep{Richings2018MNRASb}.

Another explanation of the observed scatter is that the current AGN luminosity is significantly different than the long-term average. The importance of AGN luminosity variations for outflow properties has been hinted at before. \cite{King2011MNRAS} showed that an outflow might persist for an order of magnitude longer than the AGN episode driving it; this would lead to a substantial population of `fossil' outflows, some of which have been recently identified \citep{Fluetsch2019MNRAS}. \cite{Zubovas2016MNRASb} and \cite{Zubovas2019MNRASa} showed that if the AGN is active sporadically, the outflow behaves at late times as if it was driven by the average AGN luminosity, without much variability due to sudden luminosity changes. Other authors have suggested that AGN luminosity changes might be important in establishing the high observed momentum loading factors \citep{Ishibashi2015MNRAS}, extending the AGN influence throughout the galaxy \citep{Costa2018MNRAS} and explaining the variety of observed outflow energy coupling efficiencies \citep{Harrison2018NatAs}.

This explanation is attractive because it is based on very robust observational and theoretical results. It is well known that AGN are variable on all timescales, and the relativistic winds are similarly variable as well \citep{King2003MNRASb,King2015ARAA}. Furthermore, individual AGN episodes last only $\sim 10^5$~yr \citep{King2007MNRAS, Schawinski2015MNRAS, King2015MNRAS}; over this timescale, even a very rapid outflow with $v_{\rm out} = 1000$~km~s$^{-1}$ moves only $\sim 100$~pc. Any outflows observed further than several hundred parsecs from the AGN are unlikely to have been inflated by the current AGN episode \citep{Nardini2018MNRAS, Zubovas2018MNRAS}. Furthermore, once an AGN episode begins, the outflowing material can only react to it after a time $t_{\rm react} = R_{\rm out}/v_{\rm w} \sim 3\times 10^4 R_{\rm kpc}$~yr, where $R_{\rm kpc}$ is outflow radius in kiloparsecs. This timescale is comparable to the lifetime of a single AGN episode, therefore there may be a significant lag between changes in AGN luminosity and changes in outflow properties. More importantly for the interpretation of outflow observations, the currently-observed AGN luminosity may be very different even from the luminosity that the outflow is currently reacting to, not to mention the time-averaged energy input rate. The time-averaged rate may correlate with the Eddington luminosity, producing a relation between outflow properties and SMBH mass \citep{Gonzalez2017ApJ}.

In this paper, we investigate the evolution of outflows driven by intermittent AGN episodes, with particular emphasis on the observable correlations between AGN and outflow properties. We show that this model explains the extremely small momentum- and energy-loading factors in PDS 456; we predict that the AGN in this galaxy was active $< 10\%$ of the time while the outflow has been expanding. On the other hand, the outflow in Mrk 231 is consistent with continuous driving at the present-day AGN luminosity. The outflow in IRAS F11119+3257 is intermediate between the two, suggesting non-continuous driving, albeit with a large duty cycle. We then consider the expected distribution of momentum- and energy-loading factors for a population of AGN with different duty cycles and compare that with a representative observational sample. We find that the observed and modelled distributions are broadly similar; in particular, up to $\sim 15\%$ of outflows are expected to show $\dot{p}_{\rm out} < L_{\rm AGN}/c$, although this fraction depends on both the AGN duty cycle and the host galaxy gas fraction.

We begin the paper by reviewing the salient properties of the wind feedback model, its 1D numerical implementation and the AGN luminosity prescription (Section \ref{sec:wind}). We then present the observational sample of AGN and outflow properties in Section \ref{sec:data}. We show results of modelling individual galaxies in Section \ref{sec:largescale} and the distribution of loading factors in Section \ref{sec:loading}. We discuss our results in the broader context of AGN outflows in Section \ref{sec:discuss} and summarize in Section \ref{sec:summary}.

\section{Physical and numerical outflow model} \label{sec:wind}

\subsection{Wind outflow model}

The AGN wind-driven outflow model was first proposed by \cite{King2003ApJ} and later developed mainly in \cite{King2010MNRASa} and \cite{Zubovas2012ApJ}. We refer the reader to these papers, as well as to a recent review in \cite{King2015ARAA}, for the details, and only give a brief summary of the salient points of the model.

The energy released by the AGN is communicated to the ISM via wide-angle disc winds \citep{King2003MNRASb, Nardini2015Sci}. These winds move with quasi-relativistic velocities $v_{\rm w} = \eta c \simeq 0.1 c$, where $\eta \simeq 0.1$ is the radiative efficiency of accretion. The kinetic power of the wind is $\dot{E}_{\rm w} \simeq \eta \dot{m} L_{\rm AGN} / 2 \simeq 0.05 L_{\rm AGN}$, with $\dot{m} \equiv \dot{M}_{\rm w}/\dot{M}_{\rm acc} \simeq 1$ being the mass-loading factor of the wind compared to the supermassive black hole (SMBH) accretion rate. The wind encounters the ISM, which is relatively stationary ($\sigma_{\rm turb} \ll v_{\rm w}$) and is slowed by a strong shock, which heats the wind material to temperatures of the order of $T_{\rm sh} \sim 10^{11}$~K. The subsequent evolution of the shocked wind depends on whether it cools efficiently. If the wind forms a single-temperature plasma, then cooling is efficient within several hundred parsecs of the nucleus \citep{King2003ApJ} and only the wind momentum is transferred to the ISM. If the momentum is large enough to overcome the weight of the ISM, the gas is pushed away and a large-scale outflow develops; the critical momentum rate, related to the AGN luminosity and hence the SMBH mass, establishes the $M-\sigma$ relation \citep{King2003ApJ,Murray2005ApJ}. A large-scale outflow might also develop if the shocked wind forms a two-temperature plasma, which cools inefficiently \citep{Faucher2012MNRASb}. In either case, the bubble expands adiabatically, can sweep throughout the galaxy spheroid, remove most of the gas there and quench further star formation \citep{Zubovas2012ApJ}.

This model explains the basic properties of outflows very well. In particular, it predicts that large-scale massive AGN-driven outflows should have kinetic powers $\dot{E}_{\rm out}$ equal to a few percent of $L_{\rm AGN}$ and momentum rates $\dot{p}_{\rm out}$ at least an order of magnitude greater than $L_{\rm AGN}/c$, in agreement with observations \citep{Cicone2014AA, Fiore2017AA, Gonzalez2017ApJ, Fluetsch2019MNRAS, Lutz2020AA}. However, analytical treatment of the problem is only possible for a few idealised cases, limiting the variety of galaxy properties and AGN luminosity histories that can be explored. One major limitation is that only constant AGN luminosity is tractable analytically. While the outflow is driven, it quickly reaches a quasi-steady state, with constant velocity $v_{\rm out} \sim 1000$km~s$^{-1}$ (in the rest of this paper, we adopt a scaling $v_{1000} \equiv v_{\rm out} / 1000\, {\rm km \, s}^{-1}$), mass outflow rate $\dot{M}_{\rm out}$ that can reach more than several hundred $\msun$~yr$^{-1}$ and, consequentially, the momentum and energy loading factors quoted above \citep{King2005ApJ}. When the AGN switches off, the outflow coasts for a time up to an order of magnitude longer than the duration of the AGN phase \citep{King2011MNRAS}.

If the AGN luminosity is allowed to vary significantly over time, one would expect the observed loading factors to vary as well, for two reasons. If the AGN luminosity drops suddenly, the ionization balance of the gas changes after a delay equal to the light travel time $t_{\rm l} = R_{\rm out}/c \sim 3000 R_{\rm kpc}$~yr. The decrease in ionizing flux allows some of the gas to recombine and perhaps cool down to become molecular much faster than if the AGN luminosity had been constant. Therefore, the ionized gas component of the outflow may disappear simply because the gas becomes difficult to detect, rather than because it stops moving. This interpretation is consistent with the observation that outflows in lower-luminosity AGN have a higher ratio of molecular-to-ionized mass outflow rates \citep[see Fig. 1 in][]{Fiore2017AA}. A second aspect is the dynamical response of the outflow, which cannot occur much faster than on a dynamical timescale $t_{\rm d} \simeq R_{\rm out}/v_{\rm out} \simeq 10^6 R_{\rm kpc} v_{1000}^{-1}$~yr. Naively, one might expect the observed loading factors to increase, at least for a while, as the AGN fades, if this happens on a timescale shorter than $t_{\rm d}$, and decrease as the AGN luminosity increases at the start of a new episode.

\subsection{Numerical scheme}

Here, we investigate the variations of observed momentum and energy loading factors with a 1D numerical code designed for following the outflow propagation. The code allows for any arbitrary spherically-symmetric gravitational potential and matter distribution, as long as the first and second radial derivatives of the enclosed mass, $\partial M\left(<R\right)/\partial R$ and $\partial^2 M\left(<R\right)/\partial R^2$, can be expressed analytically. The code then integrates the equation of motion for the outflow \citep[for the derivation, see][]{Zubovas2016MNRASb}:
\begin{equation} \label{eq:eom}
  \begin{split}
    \dddot{R} &= \frac{\eta L_{\rm AGN}}{M R} - \frac{2\dot{M}
      \ddot{R}}{M} - \frac{3\dot{M} \dot{R}^2}{M R} - \frac{3\dot{R}
      \ddot{R}}{R} - \frac{\ddot{M} \dot{R}}{M} \\ &
    +\frac{G}{R^2}\left[\dot{M} + \dot{M}_{\rm b} +
      \dot{M}\frac{M_{\rm b}}{M} - \frac{3}{2}\left(2M_{\rm
        b}+M\right)\frac{\dot{R}}{R}\right].
  \end{split}
\end{equation}
Here, $M$ is the mass of gas and $M_{\rm b}$ is the mass of the background (non-gaseous) matter distribution, including dark matter and stars. The time derivatives of mass are defined as $\dot{M}\equiv \dot{R}\partial M/\partial R$ and $\ddot{M} \equiv \ddot{R}\partial M/\partial R + \dot{R} ({\rm d}/{\rm d}t)\left(\partial M/\partial R\right)$. In all cases, only the relevant mass contained within the current outflow radius $R$ is considered. The first term on the right-hand side of the equation corresponds to the driving of the outflow by the AGN luminosity, the next four terms correspond to the work done by the expanding gas and the increase in outflow mass, and the remaining terms correspond to the work against gravity done while lifting the gas out of the potential well.

Each model galaxy that we investigate is composed of a halo and a bulge. The halo is assumed to have no gas and only contributes to the gravitational potential, while the bulge gas fraction is one of the free parameters of the model. Each model begins with the outflow radius and velocity set to very low values; the precise values are not important, since the outflow evolution is identical after the first few timesteps. The equation of motion (eq. \ref{eq:eom}) is integrated using a simple Eulerian integrator; other, more complicated, integration schemes produce identical results \citep{Zubovas2016MNRASb}.

The parameters we are particularly interested in are defined as follows. The outflow velocity is
\begin{equation}
    v_{\rm out} \equiv \dot{R} \equiv \iint \dddot{R} {\rm d}t {\rm d}t.
\end{equation}
The mass outflow rate is defined in a way similar to how it is usually done when analysing observations:
\begin{equation} \label{eq:mdot}
    \dot{M}_{\rm out} \equiv M_{\rm out} \frac{v_{\rm out}}{R_{\rm out}},
\end{equation}
where $M_{\rm out}$ is the total gas mass contained within $R_{\rm out}$. The momentum loading factor is
\begin{equation} \label{eq:pload}
    p_{\rm load} \equiv \frac{\dot{M}_{\rm out} v_{\rm out} c}{L_{\rm AGN}},
\end{equation}
and the energy loading factor is
\begin{equation} \label{eq:eload}
    E_{\rm load} \equiv \frac{\dot{M}_{\rm out} v_{\rm out}^2}{2L_{\rm AGN}}.
\end{equation}
In the last two cases, the instantaneous AGN luminosity is used, to mimic the properties of the system that are observationally accessible.

\subsection{AGN luminosity variation} \label{sec:lagn}

In models with varying AGN luminosity, we adopt a temporal evolution prescription based on \cite{King2007MNRAS}:
\begin{equation}
    L_{\rm AGN} = L_0 \left(1+\frac{t}{t_{\rm q}}\right)^{-19/16}.
\end{equation}
This prescription is attractive because it reproduces the observed distribution of AGN Eddington ratios; it has also been shown to be the best for keeping AGN outflow momentum and energy loading factors in the observed range \citep{Zubovas2018MNRAS}. If the AGN episode initially has $L=L_{\rm Edd}$, its duration, i.e. the time for which $L > 0.01 L_{\rm Edd}$, is $t_{\rm ep} = 48 t_{\rm q}$. Together with a recurrence timescale $t_{\rm r}$, this establishes the duty cycle of the AGN, $\delta_{\rm AGN} \equiv t_{\rm ep}/t_{\rm r}$. In this paper, we consider two values for the duty cycle: a low value $\delta_{\rm AGN} = 0.084$, approximately consistent with observational estimates of the whole galaxy population \citep{Wang2006ApJ}; and a five times larger value $\delta_{\rm AGN} = 0.42$, which represents an AGN in a prolonged period of activity, lasting several tens of Myr \citep{Yu2002MNRAS, Hopkins2005ApJ}, during which high-luminosity states are much more frequent than over the lifetime of the galaxy. In both cases, we use a recurrence time $t_{\rm r} = 10^6$~yr, while the characteristic timescales are $t_{\rm q} = 1750$~yr and $t_{\rm q} = 8750$~yr, respectively. Our chosen values of $t_{\rm q}$ correspond to AGN episode durations $t_{\rm ep} = 8.4\times10^4$~yr and $t_{\rm ep} = 4.2\times10^5$~yr, respectively, consistent with observational \citep{Schawinski2015MNRAS} and theoretical \citep{King2015MNRAS} constraints. We tested that changing both $t_{\rm q}$ and $t_{\rm rep}$ while keeping their ratio constant has almost no effect on our results, unless $t_{\rm q}$ becomes so large that the outflow sweeps through the host galaxy in a single episode, but this is unrealistic.

\section{Outflow data} \label{sec:data}

The AGN sample considered in this work is largely drawn from the recent collections of \citet{Gonzalez2017ApJ}, \citet{Fluetsch2019MNRAS} and \citet{Lutz2020AA}. For many objects, the large-scale outflow is also detected in the neutral and/or ionized phase (e.g. \citealt{Rupke2017ApJ}), with the molecular component that is usually dominant by $\sim$1--2 orders of magnitude over the ionized one \citep[see also][]{RobertsBorsani2020MNRAS}.\footnote{There is a possible trend towards equi-partition between the molecular and ionized phases at the highest AGN luminosities ($L_{\rm AGN} \ga 10^{47}$ erg s$^{-1}$; e.g. \citealt{Fiore2017AA}), but the typical AGN luminosity in our sample is much lower.} Whenever the cold gas phase is probed in both CO and OH, some discrepancy exists between mass, momentum, and kinetic energy outflow rates based on these two tracers. This difference is mostly due to the smaller (by factors of a few) radii inferred from the OH features, while masses and (especially) velocities are in good agreement (see also Fig. 6 of \citealt{Lutz2020AA}). The origin of this effect is not completely clear, so here we stick to CO-based radii, which are directly measured through imaging data. 

Even limited to CO studies, however, there is no accepted convention on the definition of the key outflow properties. For instance, once the association of broad emission lines with the outflow is established (as opposed to a regular velocity field like a rotating disc), the flux is variously integrated over the entire profile or only over its wings, and the outflow velocity can be either identified with the bulk (peak) velocity $v_{\rm broad}$ or with the maximum velocity, $v_{\rm max} = v_{\rm broad}+2\sigma$, the latter believed to be a proxy of the true (deprojected) gas velocity. Hence, not only the mass (obtained from the flux) but also the dynamical properties, whereby the momentum and energy rates scale as the second and third power of $v_{\rm out}$, can vary by factors of several simply based on the starting assumptions. Here we therefore give priority to the estimates of \citet{Lutz2020AA}, who revised the literature data with a uniform prescription, for which the outflow velocity is the sum of the net shift of the broad component compared to the systemic velocity and its half width at a tenth of its peak, and the wings are defined such as the broad component accounts for more than 50\% of the total line emission. 

When computed as in eq. \ref{eq:mdot}, the mass outflow rate corresponds to an isothermal density profile with $R_{\rm out} \gg R_{\rm in}$ (e.g. \citealt{Rupke2005ApJS}) and it assumes a constant value in both radius and time. This is clearly an oversimplification, but it represents a convenient and sensible benchmark to make the properties of outflows as obtained in different studies uniform. Several works adopt instead a constant density in the outflow, as this has the advantage that the mass outflow rate does not depend on the exact geometrical structure (i.e., the bi-cone opening angle; e.g. \citealt{Maiolino2012MNRAS,Fiore2017AA}). This assumption brings a multiplicative factor in the expression above of just 3, yet it is less appropriate for a comparison with our model, also in terms of the implied outflow history (see the discussion in \citealt{Lutz2020AA}). 

Another major source of uncertainty in the characterization of large-scale outflows resides in the conversion between the observed CO line luminosity ($L^\prime_\rmn{CO}$) and the total molecular mass  ($M_{\rmn H_2}=M_{\rm out}$), which is usually implemented through the mass-to-light ratio factor $\alpha_\rmn{CO}$ (e.g. \citealt{Bolatto2013ARAA}). The typical range of $\alpha_\rmn{CO}$ is 0.34--4.3 $M_{\sun}$ (K km s$^{-1}$ pc$^2$)$^{-1}$, where the lower limit corresponds to the optically-thin case, while the upper limit coincides with the Galactic value.\footnote{For simplicity, the units of $\alpha_\rmn{CO}$ are omitted hereafter.} These values are generally derived for the quiescent gas, whose physical conditions are not necessarily representative of the outflow components, for which few constraints are available instead. In some cases, CO emission lines from higher order transitions than $J$=1--0 are probed. Line luminosity ratios $r_{J_\rmn{up}1}$ must then be introduced. However, for low-excitation transitions like $J$=3--2 and $J$=2--1 considered here, $r_{J_\rmn{up}1} \approx 1$ (e.g. \citealt{Papadopoulos2012ApJ}), hence this can be seen as an additional source of uncertainty on $\alpha_\rmn{CO}$ itself. In this work, we adopt $\alpha_\rmn{CO}=0.8$, as per the rotating nuclear discs of ULIRGs \citep{Downes1998ApJ}, which represent a substantial fraction of the literature samples. Although values as low as $\alpha_\rmn{CO}=0.5$ are used in some works, our choice is still rather conservative if compared, for instance, to $\alpha_\rmn{CO}\sim 2.1$, as measured in the outflow of NGC 6240 \citep{Cicone2018ApJ}.

In our estimates of the outflowing molecular mass and of the dependent dynamical properties, we account for the scatter of $\alpha_\rmn{CO}$ in individual objects by assuming an uncertainty of 0.3 dex on the total molecular masses of each component. In any case, we expect the choice of a common $\alpha_\rmn{CO}$ not to have any global effects on our results, as these will cancel out over a sufficiently large sample. Moreover, while the presence of residual, $\alpha_\rmn{CO}$-related systematics cannot be entirely ruled out, we are confident that these would mostly shift any correlations with the AGN properties by acting on their normalizations, rather than inducing spurious correlations or concealing real ones.

\clearpage
\onecolumn

\begin{table*}
\begin{ThreePartTable}
\scriptsize 
\begin{TableNotes}
\item[] \textit{Notes.} Source redshifts were retrieved from the NASA/IPAC Extragalactic Database, \url{http://ned.ipac.caltech.edu/}. AGN luminosities are based on different multiwavelength indicators (such as IR colours and [O\,\textsc{iv}]\,25.9 $\mu$m, 6 $\mu$m, 5100 \AA, [O\,\textsc{iii}]\,5007 \AA, 1350 \AA, and 2--10 keV luminosities]), and have uncertainties of 0.1--0.4 dex. Outflow properties were adapted from the original references as described in Section \ref{sec:data}. For sources with spatially distinct components, (a) is close to the nucleus and (b) is farther out.
\end{TableNotes}
\begin{longtable}{l@{\hspace{25pt}}c@{\hspace{15pt}}c@{\hspace{15pt}}c@{\hspace{15pt}}c@{\hspace{15pt}}c@{\hspace{15pt}}l}
\caption{Molecular outflow properties of the sources considered in our analysis.} \\
\label{ts} \\
\hline \\[-2.5ex]
Source & $z$ & $\log\,L_{\rm AGN}$ & $\log\,\dot{M}_{\rm out}$ & 
$\log\,\dot{P}_{\rm out}$ & $\log\,\dot{E}_{\rm out}$ & Reference \\
       &       & (erg s$^{-1}$)      & ($M_\odot$ yr$^{-1}$)     & 
($L_{\rm AGN}/c$)         & ($L_{\rm AGN}$)            &          \\
\hline \\[-2ex]
NGC\,253            & 0.001 & 42.3      & 3.4$^{+3.7}_{-1.8}$    & 
1.2$^{+0.4}_{-0.3}$  & $-2.8^{+0.4}_{-0.4}$ & \citet{Bolatto2013Nat} \\[0.7ex]
III\,Zw\,35         & 0.027 & 42.7      & 57$^{+60}_{-30}$       & 
2.9$^{+0.4}_{-0.5}$  & $-0.4^{+0.4}_{-0.5}$ & \citet{Lutz2020AA} \\[0.7ex]
PG\,0157+001        & 0.163 & 45.7      & 63$^{+68}_{-33}$       & 
$-0.1^{+0.4}_{-0.3}$ & $-3.2^{+0.3}_{-0.4}$ & \citet{Lutz2020AA} \\[0.7ex]
NGC\,1068           & 0.004 & 45.4      & 14$^{+15}_{-7}$        & 
$-1.0^{+0.3}_{-0.4}$ & $-4.8^{+0.4}_{-0.4}$ & \citet{GarciaB2014AA} \\[0.7ex]
NGC\,1266           & 0.007 & $<\,$42.2 & 39$^{+42}_{-21}$       & 
$>\,$2.8             & $>\,$$-0.5$          & \citet{Lutz2020AA} \\[0.7ex]
NGC\,1377           & 0.006 & 42.2      & 6.6$^{+7.1}_{-3.5}$    & 
2.0$^{+0.5}_{-0.4}$  & $-1.6^{+0.5}_{-0.5}$ & \citet{Aalto2012AA} \\[0.7ex]
NGC\,1433           & 0.004 & 42.1      & 0.9$^{+1.1}_{-0.5}$    & 
1.2$^{+0.4}_{-0.4}$  & $-2.6^{+0.5}_{-0.5}$ & \citet{Combes2013AA} \\[0.7ex]
NGC\,1614           & 0.016 & 44.0      & 19$^{+20}_{-10}$       & 
1.1$^{+0.4}_{-0.3}$  & $-2.1^{+0.4}_{-0.4}$ & \citet{GarciaB2015AA} \\[0.7ex]
NGC\,1808           & 0.003 & 42.8      & 3.4$^{+3.6}_{-1.7}$    & 
1.0$^{+0.5}_{-0.5}$  & $-2.8^{+0.6}_{-0.5}$ & \citet{Salak2016ApJ} \\[0.7ex]
IRAS\,05083+7936    & 0.054 & 44.2      & 79$^{+86}_{-42}$       & 
1.9$^{+0.4}_{-0.4}$  & $-1.0^{+0.5}_{-0.4}$ & \citet{Lutz2020AA} \\[0.7ex]
IRAS\,05189$-$2524  & 0.043 & 45.6      & 37$^{+39}_{-19}$       & 
$-0.1^{+0.3}_{-0.4}$ & $-3.3^{+0.4}_{-0.3}$ & \citet{Lutz2020AA} \\[0.7ex]
NGC\,2146           & 0.003 & 42.9      & 10$^{+10}_{-6}$        & 
1.7$^{+0.3}_{-0.4}$  & $-1.8^{+0.4}_{-0.4}$ & \citet{Tsai2009PASJ} \\[0.7ex]
NGC\,2623\,(a)        & 0.019 & 44.2      & 11$^{+13}_{-6}$        & 
0.9$^{+0.3}_{-0.3}$  & $-2.1^{+0.3}_{-0.4}$ & \citet{Lutz2020AA} \\[0.7ex]
NGC\,2623\,(b)        &       &           & 0.4$^{+0.4}_{-0.2}$    & 
$-0.8^{+0.4}_{-0.3}$ & $-4.0^{+0.4}_{-0.3}$ & \citet{Lutz2020AA} \\[0.7ex]
IRAS\,08572+3915    & 0.058 & 45.7      & 345$^{+376}_{-176}$    & 
1.1$^{+0.3}_{-0.3}$  & $-1.8^{+0.5}_{-0.7}$ & \citet{Cicone2014AA} \\[0.7ex]
IRAS\,09111$-$1007W & 0.054 & $<\,$43.5 & 64$^{+70}_{-33}$       &
$>\,$2.0             & $>\,$$-1.3$          & \citet{Lutz2020AA} \\[0.7ex]
M\,82               & 0.001 & 41.5      & 6.7$^{+7.0}_{-3.5}$    & 
2.6$^{+0.4}_{-0.4 }$ & $-1.2^{+0.4}_{-0.4}$ & \citet{Walter2002ApJ} \\[0.7ex]
NGC\,3256           & 0.009 & 44.0      & 50$^{+39}_{-22}$       & 
1.8$^{+0.3}_{-0.3}$  & $-1.2^{+0.3}_{-0.3}$ & \citet{Sakamoto2014ApJ} \\[0.7ex]
IRAS\,10565+2448    & 0.043 & 44.8      & 95$^{+99}_{-49}$       & 
1.1$^{+0.5}_{-0.4}$  & $-2.0^{+0.5}_{-0.5}$ & \citet{Cicone2014AA} \\[0.7ex]
IRAS\,11119+3257\,(a) & 0.189 & 46.0      & 203$^{+222}_{-108}$    & 
0.6$^{+0.4}_{-0.4}$  & $-2.2^{+0.5}_{-0.5}$ & \citet{Tombesi2015Natur} \\[0.7ex]
IRAS\,11119+3257\,(b) &       &           & 81$^{+92}_{-42}$       & 
0.2$^{+0.4}_{-0.4}$  & $-2.6^{+0.5}_{-0.5}$ & \citet{Veilleux2017ApJ} \\[0.7ex]
NGC\,3628           & 0.003 & 41.6      & 1.3$^{+1.4}_{-0.7}$    & 
1.8$^{+0.6}_{-0.5}$  & $-2.0^{+0.5}_{-0.5}$ & \citet{Tsai2012ApJ} \\[0.7ex]
ESO\,320--G030      & 0.011 & 41.8      & 10$^{+3}_{-1}$         & 
3.0$^{+1.1}_{-0.4}$  & $-0.3^{+1.1}_{-0.4}$ & \citet{PereiraS2016AA} \\[0.7ex]
IRAS\,12112+0305    & 0.073 & 44.8      & 198$^{+198}_{-96}$     &
1.4$^{+0.7}_{-0.4}$  & $-1.7^{+0.7}_{-0.4}$ & \citet{PereiraS2018AA} \\[0.7ex]
IRAS\,12224$-$0624  & 0.026 & 41.7      & 23$^{+24}_{-12}$       & 
3.6$^{+0.5}_{-0.4}$  & 0.6$^{+0.5}_{-0.5}$  & \citet{Lutz2020AA} \\[0.7ex]
NGC\,4418           & 0.007 & 44.4      & 9.1$^{+9.5}_{-4.6}$    & 
0.3$^{+0.6}_{-0.5}$  & $-2.9^{+0.5}_{-0.6}$ & \citet{Lutz2020AA} \\[0.7ex]
Mrk\,231            & 0.042 & 46.0      & 330$^{+375}_{-169}$    &
0.7$^{+0.4}_{-0.4}$  & $-2.2^{+0.4}_{-0.4}$ & \citet{Cicone2012AA} \\[0.7ex]
IRAS\,13120$-$5453  & 0.031 & 43.8      & 134$^{+147}_{-68}$    & 
2.1$^{+0.4}_{-0.4}$  & $-1.2^{+0.4}_{-0.4}$ & \citet{Lutz2020AA} \\[0.7ex]
M\,51               & 0.002 & 43.2      & 1.9$^{+2.1}_{-1.0}$    & 
0.3$^{+0.5}_{-0.4}$  & $-3.4^{+0.5}_{-0.6}$ & \citet{Querejeta2016AA} \\[0.7ex]
Mrk\,273            & 0.038 & 45.5      & 190$^{+197}_{-97}$     & 
1.1$^{+0.5}_{-0.4}$  & $-2.1^{+0.4}_{-0.4}$ & \citet{Cicone2014AA} \\[0.7ex]
4C\,+12.50          & 0.122 & 45.6      & 128$^{+140}_{-65}$     & 
0.6$^{+0.4}_{-0.4}$  & $-1.8^{+0.4}_{-0.4}$ & \citet{Dasyra2011AA} \\[0.7ex]
J135646.10+102609.0 & 0.123 & 45.7      & 109$^{+119}_{-56}$     & 
0.3$^{+0.5}_{-0.4}$  & $-2.7^{+0.4}_{-0.5}$ & \citet{Sun2014ApJ} \\[0.7ex]
Circinus Galaxy     & 0.001 & 43.4      & 1.3$^{+1.4}_{-0.7}$    & 
0.1$^{+0.4}_{-0.3}$  & $-3.5^{+0.4}_{-0.3}$ & \citet{Zschaechner2016ApJ} \\[0.7ex]
IRAS\,14348$-$1447  & 0.083 & 45.1      & 199$^{+154}_{-85}$     & 
1.1$^{+0.5}_{-0.4}$  & $-2.1^{+0.6}_{-0.4}$ & \citet{PereiraS2018AA} \\[0.7ex]
IRAS\,14378$-$3651  & 0.068 & $<\,$44.0 & 208$^{+153}_{-88}$     & 
$>\,$2.0             & $>\,$$-1.0$          & \citet{Gonzalez2017ApJ} \\[0.7ex]
Arp\,220            & 0.018 & 45.0      & 52$^{+43}_{-22}$       & 
0.7$^{+0.6}_{-0.3}$  & $-2.4^{+0.5}_{-0.4}$ & \citet{Barcos2018ApJ} \\[0.7ex]
Mrk\,876            & 0.129 & 45.8      & 738$^{+769}_{-377}$    & 
1.4$^{+0.3}_{-0.3}$  & $-1.4^{+0.3}_{-0.4}$ & \citet{Lutz2020AA} \\[0.7ex]
NGC\,6240           & 0.024 & 44.9      & 262$^{+298}_{-140}$    & 
1.4$^{+0.4}_{-0.4}$  & $-1.8^{+0.5}_{-0.4}$ & \citet{Feruglio2013AA} \\[0.7ex]
IRAS\,17020+4544    & 0.060 & 45.1      & 322$^{+336}_{-164}$    & 
1.9$^{+0.5}_{-0.5}$  & $-0.7^{+0.5}_{-0.4}$ & \citet{Lutz2020AA} \\[0.7ex]
IRAS\,17208$-$0014  & 0.043 & 45.0      & 112$^{+121}_{-56}$     & 
1.4$^{+0.9}_{-0.4}$  & $-1.3^{+0.8}_{-0.5}$ & \citet{Lutz2020AA} \\[0.7ex]
PDS\,456\,(a)         & 0.184 & 47.1      & 91$^{+74}_{-40}$       & 
$-1.0^{+0.4}_{-0.3}$ & $-3.9^{+0.5}_{-0.3}$ & \citet{Bischetti2019AA} \\[0.7ex]
PDS\,456\,(b)         &       &           & 21$^{+11}_{-6}$        & 
$-1.6^{+0.4}_{-0.3}$ & $-4.5^{+0.4}_{-0.3}$ & \citet{Bischetti2019AA} \\[0.7ex]
NGC\,6764           & 0.008 & 42.7      & 0.02$^{+0.02}_{-0.01}$ &
$-1.7^{+0.4}_{-0.4}$ & $-6.1^{+0.4}_{-0.4}$ & \citet{Leon2007AA} \\[0.7ex]
IRAS\,20100$-$4156  & 0.130 & 45.5      & 689$^{+751}_{-359}$    & 
1.6$^{+0.3}_{-0.4}$  & $-1.3^{+0.4}_{-0.3}$ & \citet{Gowardhan2018ApJ} \\[0.7ex]
IC\,5063            & 0.011 & 44.6      & 13$^{+13}_{-7}$       & 
0.3$^{+0.3}_{-0.3}$  & $-3.0^{+0.3}_{-0.4}$ & \citet{Morganti2013AA} \\[0.7ex]
IRAS\,20551$-$4250  & 0.043 & 45.1      & 67$^{+74}_{-35}$       & 
0.7$^{+0.4}_{-0.3}$  & $-2.3^{+0.3}_{-0.4}$ & \citet{Lutz2020AA} \\[0.7ex]
IRAS\,22491$-$1808  & 0.078 & 44.9      & 91$^{+94}_{-47}$       & 
0.8$^{+0.4}_{-0.4}$  & $-2.5^{+0.4}_{-0.3}$ & \citet{PereiraS2018AA} \\[0.7ex]
IRAS\,23060+0505    & 0.173 & 45.9      & 361$^{+362}_{-180}$    & 
0.6$^{+0.3}_{-0.3}$  & $-2.4^{+0.3}_{-0.3}$ & \citet{Lutz2020AA} \\[0.7ex]
IRAS\,23365+3604    & 0.064 & 44.6      & 51$^{+54}_{-27}$       &
1.1$^{+0.4}_{-0.4}$  & $-2.1^{+0.5}_{-0.4}$ & \citet{Cicone2014AA} \\[0.7ex]
\hline \\
\insertTableNotes
\end{longtable}
\end{ThreePartTable}
\end{table*}

\clearpage
\twocolumn

\begin{table*}
\begin{center}
\begin{tabular}{ |l|c|c|c|c } 
 \hline
 Parameter & Mrk 231 & PDS 456 & IRAS F11119+3257 & Seyfert\\ 
 \hline
 Virial mass $\left(M_{\rm vir}\right)$ & $1.18\times10^{13} \,\msun$ & $1.26\times10^{12} \,\msun$ & $10^{15} \,\msun$ & $2\times10^{11} \,\msun$ \\ 
 Virial radius $\left(r_{\rm vir}\right)$ & $662$~kpc & $235$~kpc & $2905$~kpc & $180.5$~kpc\\ 
 Concentration $\left(c\right)$           & $10$ & $10$ & $10$ & $10$\\ 
 Bulge radius $\left(r_{\rm bulge}\right)$ & $2$~kpc & $5$~kpc & $15$~kpc & $2$~kpc \\ 
 Bulge mass $\left(M_{\rm bulge}\right)$ & $5.4\times10^{10} \,\msun$ & $3.8\times10^{10} \,\msun$ & $10^{11} \,\msun$ & $8\times10^{9} \,\msun$ \\ 
 \hline
 SMBH mass $\left(M_{\rm SMBH}\right)$ & $1.8\times10^8 \,\msun$ & $1.4\times10^9 \,\msun$ & $1.0\times10^8 \,\msun$ & $5\times10^6 \,\msun$ \\ 
 AGN luminosity (Eddington ratio): & & & & \\
 Continuous $\left(L_{\rm AGN, c} \left(f_{\rm c}\right)\right)$ & $9.0\times10^{45}$ erg s$^{-1} \,(0.4)$ & $1.2\times10^{47}$ erg s$^{-1} \,(0.7)$ & $1.1\times10^{46}$ erg s$^{-1} \,(0.9)$ & n/a\\ 
 Initial $\left(L_{\rm 0, r} \left(f_{\rm 0, r}\right)\right)$ & \del{$6.2 \times 10^{46}$ erg s$^{-1} \,(2)$} n/a & $1.7\times10^{47}$ erg s$^{-1} \,(1)$ & $2.9\times10^{46}$ erg s$^{-1} \,(2.3)$ & $3.3 \times 10^{44}$ erg s$^{-1} \,(0.5)$ \\ 
 Duty cycle $\left(\delta_{\rm AGN}\right)$ & n/a & $0.084$ & $0.42$ & $0.084$ \\ 
 \hline
\end{tabular}
\caption{Parameters of the Mrk 231, PDS 456, IRAS F11119+3257 and `generic Seyfert' galaxy models. Virial masses and radii are estimated from rotational velocities. Bulge radii are set equal to the radius within which gas mass estimates are provided in the literature, and bulge masses are calculated from observationally determined velocity dispersions. SMBH masses were obtained from the H$\beta$ line width and the de-reddened 5100 \AA\ luminosity, applying the radius--luminosity relation of \citet{Bentz2013ApJ} with a virial factor of 4.3. The data were retrieved from \citet{Zheng2002AJ} for Mrk 231 and IRAS F11119, and from \citet{Torres1997ApJ} and \citet{Simpson1999MNRAS} for PDS 456. AGN luminosities were obtained from the different IR diagnostics of \citet{Veilleux2009ApJS} for Mrk 231 and IRAS F11119, and from the de-reddened 1350 \AA flux \citep{Hamann2018MNRAS} with bolometric correction by \citet{Richards2006ApJS} for PDS 456.}
\label{table:params}
\end{center}
\end{table*}

When not explicitly reported, we use uncertainties of 10 and 20 per cent, respectively, on outflow velocities and radial distances. Uncertainties on outflow properties are propagated through Monte-Carlo simulations assuming normal or lognormal distributions for the quantities involved. 
The resulting median and 16$^{\rm th}$--84$^{\rm th}$ percentiles are assumed as best value and confidence range. Any correction for projection effects is neglected, even for the few objects where tentative geometrical information is available. Finally, we treat the advanced AGN mergers with a close pair of nuclei as single systems, since the luminosity of each AGN cannot be disentangled.

The salient AGN and outflow properties of the 43 objects in our sample are given in Table \ref{ts}.

\section{Results: individual objects} \label{sec:largescale}

Before investigating how well our model reproduces the statistical properties of the whole observational sample, we consider the propagation of outflows in three model galaxies, corresponding to Mrk 231, PDS 456 and IRAS F11119+3257. These objects were chosen for several reasons. First of all, they are all type 1 quasars with disturbed yet compact morphology, likely in the post-coalescence stage, so the spherical symmetry of our models is not too crude an approximation; this would not be the case in, say, the double nuclei of early-stage mergers or Seyfert spiral galaxies. Secondly, all three objects have been reasonably well studied and show evidence of both a small-scale quasi-relativistic wind and one or more large-scale outflows. Finally, these objects help us very clearly illustrate the difference between continuous and intermittent AGN driving.

The salient properties of those galaxies that we used to set up the models are given in Table \ref{table:params}. For each galaxy, we test models with different gas fractions in the bulge, giving $M_{\rm g} = f_{\rm g} M_{\rm bulge}$. We assume the bulge density profile to be isothermal and the rest of the mass to be distributed in a halo with an NFW \citep{Navarro1997ApJ} profile. We also assume that the bulge extends out to the most distant outflow radius, because the gas mass estimates are typically available only at that radius. Therefore, our results should only be considered physical out to the assumed bulge radius; in all subsequent plots, outflow propagation within the bulge is marked with thicker lines than outside it. Within our models, outflow propagation inside the bulge is quite insensitive to the assumed halo density profile, virial mass and virial radius of each galaxy, as long as $r_{\rm vir} \gg R_{\rm out}$. Each model was integrated until the outflow reached $R_{\rm max} = 25$~kpc, which typically took several tens of Myr. Expansion within the physically-constrained bulge region typically takes a few Myr in each model.

\begin{figure}
	\includegraphics[width=\columnwidth]{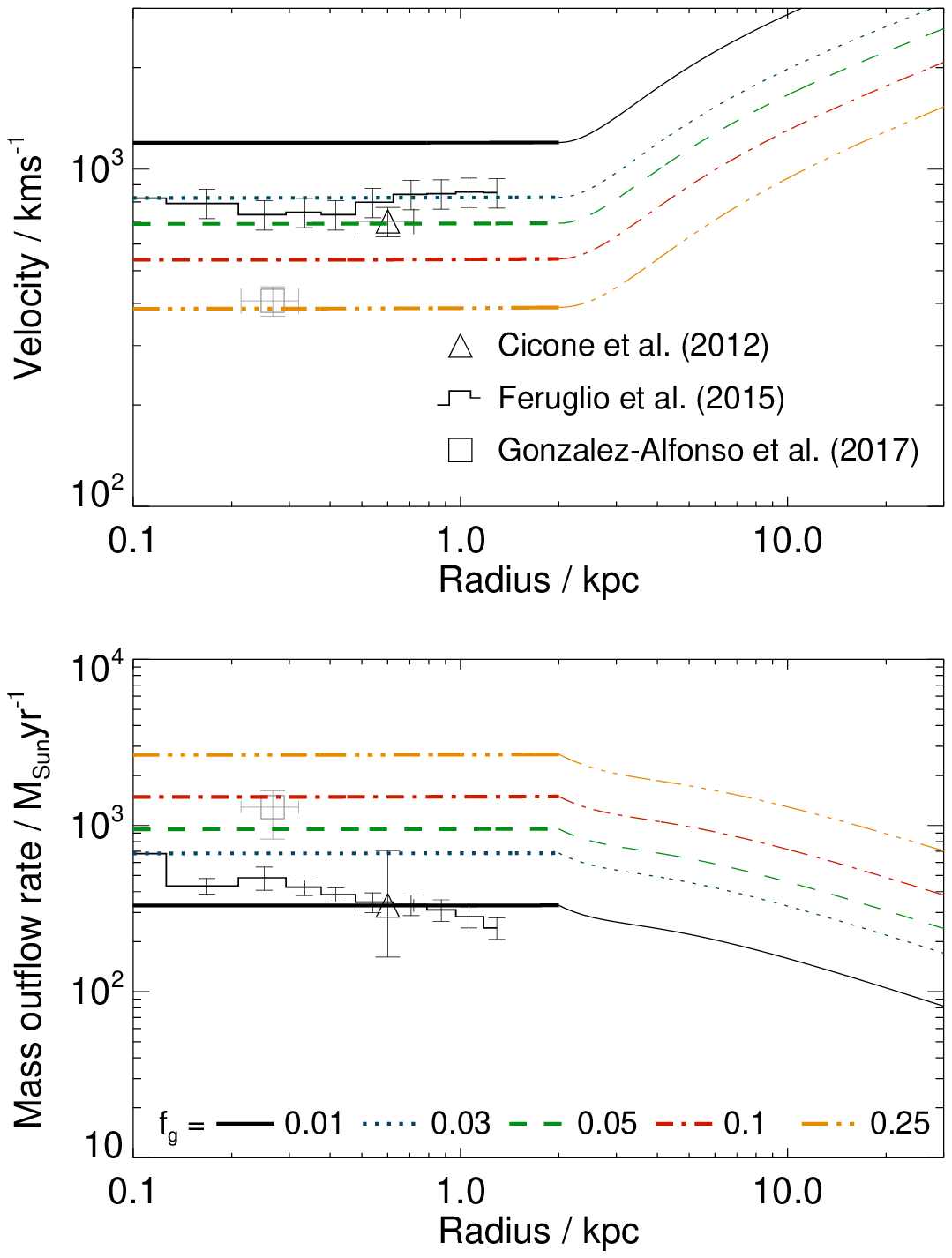}
    \caption{Outflow evolution in Mrk 231 with a continuous AGN energy injection, for models with different gas fractions. Observed outflow properties marked with triangles \citep[CO]{Cicone2012AA}, histograms \citep[CO]{Feruglio2015AA} and squares \citep[OH]{Gonzalez2017ApJ}.}
    \label{fig:Mrk_largescale}
\end{figure}

Figure \ref{fig:Mrk_largescale} shows the radial profiles of velocity and mass outflow rate in the Mrk 231 model, with the AGN active continuously. The five lines correspond to models with different gas fractions, from gas poor $f_{\rm g} = 0.01$ (black solid) to gas rich $f_{\rm g} = 0.25$ (orange triple-dot-dashed). We also show the best current observational estimates of outflow properties: CO data from \cite[triangles]{Cicone2012AA} and from \cite[histogram, rescaled to $\alpha_\rmn{CO}=0.8$]{Feruglio2015AA}, as well as OH data from \cite[grey squares]{Gonzalez2017ApJ}.\del{We show a conservative error of $20\%$ in the radius estimates from \cite{Cicone2012AA} and \cite{Gonzalez2017ApJ}; otherwise the errors are the same as given in the three papers.} The notable difference between OH and CO data comes mainly from differences in radius estimates (see Section \ref{sec:data}), and we generally consider the more conservative CO data in the following discussion. Note, however, that it is possible that OH data traces a denser, and hence slower, outflow in this object.

The almost constant outflow velocity with radius derived by \cite{Feruglio2015AA}, which agrees with the earlier result \citep{Cicone2012AA}, is well explained by our model, assuming a rather low gas fraction ($0.02 < f_{\rm g} < 0.05$). The inferred molecular gas content \citep[$M_{{\rm H}_2} \simeq 5\times10^9 \msun$;][]{Cicone2012AA} corresponds to a gas fraction $f_{\rm g} \sim 0.09$, somewhat higher than our preferred values. However, the mass estimate depends sensitively on the assumed conversion factor of CO to H$_2$, with an uncertainty of at least a factor of two, which brings our model in line with the observational estimate. The mass outflow rate predicted by our models is marginally higher than observed: $\dot{M}_{\rm model} \sim 500-1000 \msun$~yr$^{-1}$, compared with $\dot{M}_{\rm obs} \sim 150-700 \msun$~yr$^{-1}$. However, one should keep in mind that our models assume perfectly spherically symmetric outflows; if the outflow only encompasses a solid angle of $2\pi$ as seen from the nucleus, the expected mass outflow rate drops by a factor two and comes well in line with observational estimates. We also note that \cite{Feruglio2015AA} find the mass outflow rate decreasing slightly with radius, a result which our models do not replicate. Two possible solutions to the discrepancy are a different gas density distribution (however, this would result in the outflow velocity increasing with radius) and/or the outflow solid angle decreasing with radius. This latter possibility seems rather natural if some of the outflowing material slows down and never makes it out of the galaxy, as recently suggested by \cite{Fluetsch2019MNRAS}.

The momentum- and energy-loading factors of the modelled outflows fall in the range $8 < p_{\rm load} < 20$ and $0.012 < E_{\rm load} < 0.015$, close to the observationally-derived values $p_{\rm load,obs} \simeq 5$ and $E_{\rm load, obs} \sim 0.006$ \citep{Cicone2012AA,Feruglio2015AA}. The outflow in Mrk 231 also has a substantial neutral component \citep{Rupke2011ApJ, Rupke2017ApJ}, so the total momentum and energy loading factors are consistent with model predictions. In addition, the outflow might be non-spherical and some of its material might slow down as it moves away from the nucleus (see preceding paragraph); these two effects would result in lower loading factors than in the idealized case.

\begin{figure}
	\includegraphics[width=\columnwidth]{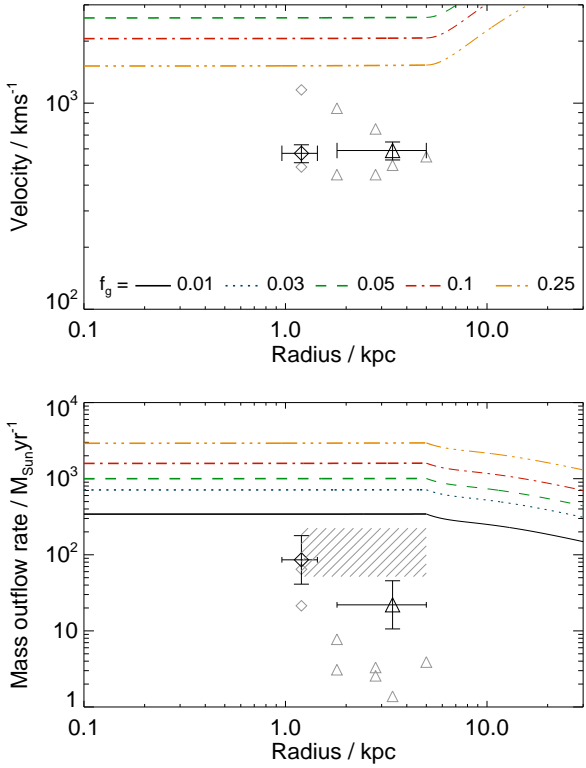}
    \caption{Same as Figure \ref{fig:Mrk_largescale}, but for the model of PDS 456. Points show observational data \citep{Bischetti2019AA}: diamond represents the central component, triangle corresponds to the extended outflow. Fainter points show sub-components. The shaded region in the bottom plot represents the sum of mass outflow rates of both central an extended components.}
    \label{fig:PDS_largescale}
\end{figure}

Outflow propagation in PDS 456 shows a marked contrast with the Mrk 231 model. In Figure \ref{fig:PDS_largescale} we show the properties of an outflow in this system driven by a constant-luminosity AGN. The triangle and diamond data points correspond to the ``extended'' and ``central'' outflow components identified by \cite{Bischetti2019AA}, with grey points showing smaller sub-components and the shaded region in the bottom plot showing the total mass outflow rate. We show a $20\%$ radius error for the central component, while all other errors are as given by \cite{Bischetti2019AA}, with appropriate rescalings (a constant density was assumed in the original analysis for the central outflow). The very low observed values of momentum and energy loading factors led these authors to suggest that the outflow in PDS 456 might be driven by the AGN momentum, rather than energy input. Our model results also clearly show that an outflow driven by the energy of a continuous AGN episode has properties inconsistent with the observed data: the model outflow is either much faster than the observed one, or carries much more mass, or both. One way of reconciling these results with observations is by assuming that the molecular gas comprises only a small fraction of the total outflow. However, in order to get the right outflow velocity, the fraction of the bulge mass participating in the outflow should be $f_{\rm g} > 0.25$, i.e. $M_{\rm gas}\left(<5 {\rm kpc}\right) > 9.5\times10^9 \, \msun$, which is at least a factor of 40 greater than the observed outflowing molecular gas mass \citep{Bischetti2019AA}. In general, when molecular gas is observed, it tends to dominate the mass budget of the outflow (see Section \ref{sec:data}). Therefore we are confident that the observed molecular outflow comprises a significant, and most probably dominant, part of the total outflow.

\begin{figure}
	\includegraphics[width=\columnwidth]{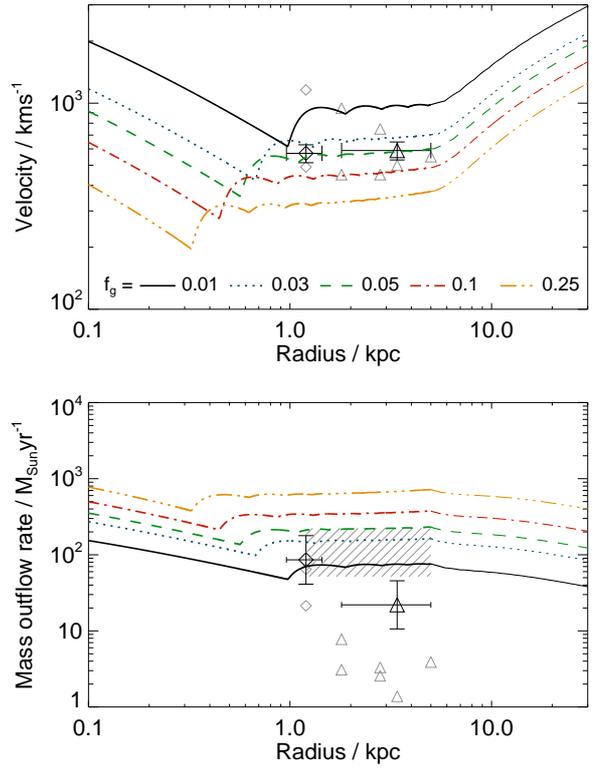}
    \caption{Same as Figure \ref{fig:PDS_largescale}, but for a varying AGN luminosity history with $\delta_{\rm AGN} = 0.084$.}
    \label{fig:PDS_rep_largescale}
\end{figure}

Another way to reconcile observations and model results is to consider a non-continuous AGN luminosity history. Figure \ref{fig:PDS_rep_largescale} shows the results of such a simulation, where we adopt the luminosity history as described in Section \ref{sec:lagn} with a duty cycle $\delta_{\rm AGN} = 0.084$. Both outflow velocities and mass flow rates decrease significantly compared to the continuous-AGN model, and agree quite well with observed data, assuming $0.03 < f_{\rm g} < 0.05$. After a few AGN episodes, the variation of AGN luminosity no longer has a noticeable effect on the outflow velocity. This means that the values of momentum and energy loading factors vary significantly with each AGN episode, being inversely proportional to $L_{\rm AGN}$ \citep[see also][]{Zubovas2018MNRAS}. There are periods of time when the AGN can be observed at its present-day luminosity $L_{\rm AGN} = 1.2\times10^{47}$~erg~s$^{-1}$ simultaneously with the observed outflow properties. The momentum and energy loading factors at those times are $\dot{p}_{\rm out} < L_{\rm AGN}/c$ and $\dot{E}_{\rm out} < 10^{-3} L_{\rm AGN}$, again consistent with observations \citep{Bischetti2019AA}.

\begin{figure}
	\includegraphics[width=\columnwidth]{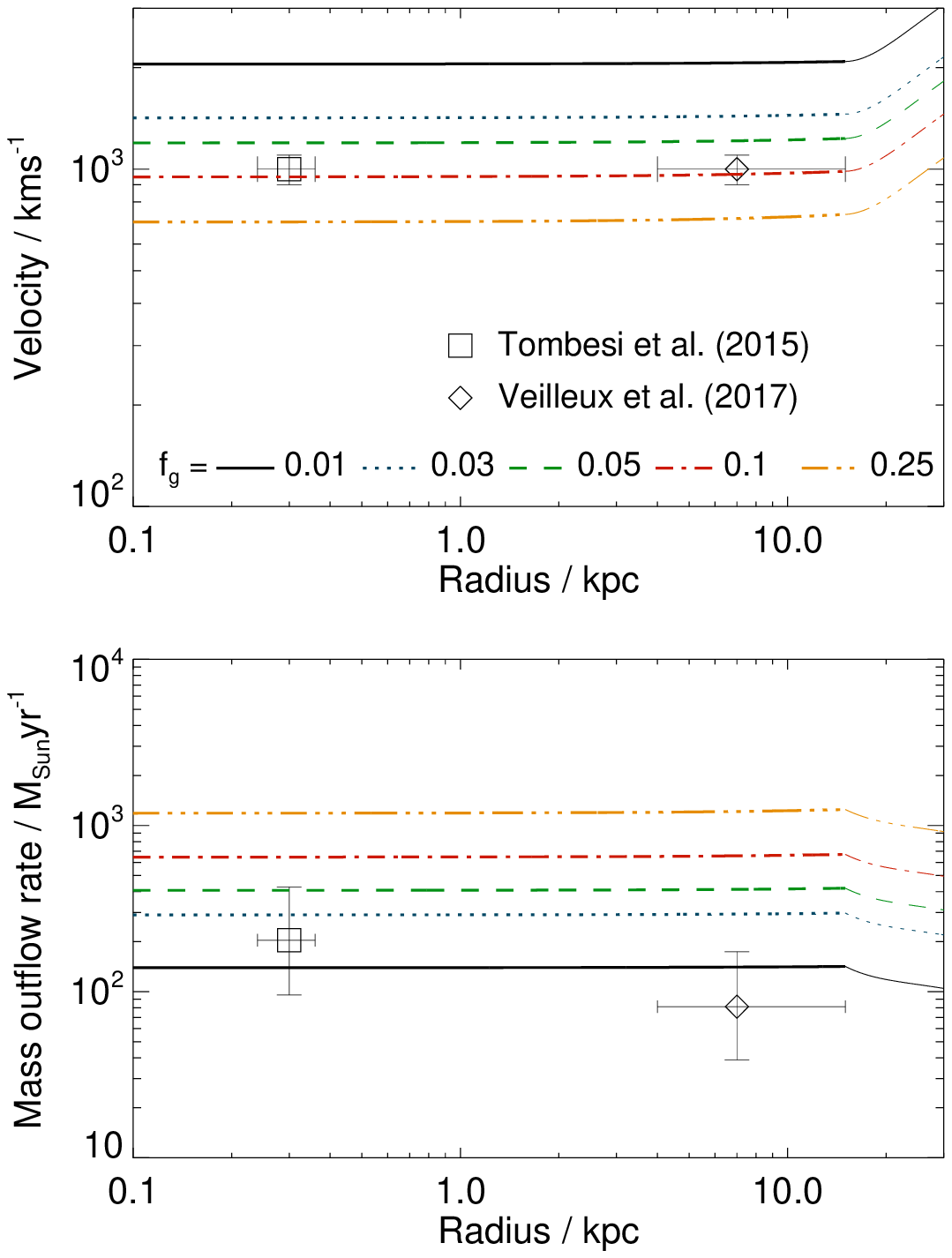}
    \caption{Same as Figure \ref{fig:Mrk_largescale}, but for the model of IRAS F11119+3257. Points represent outflow data from \citep[squares]{Tombesi2015Natur} and \citep[diamonds]{Veilleux2017ApJ}.}
    \label{fig:F11119_largescale}
\end{figure}

\begin{figure}
	\includegraphics[width=\columnwidth]{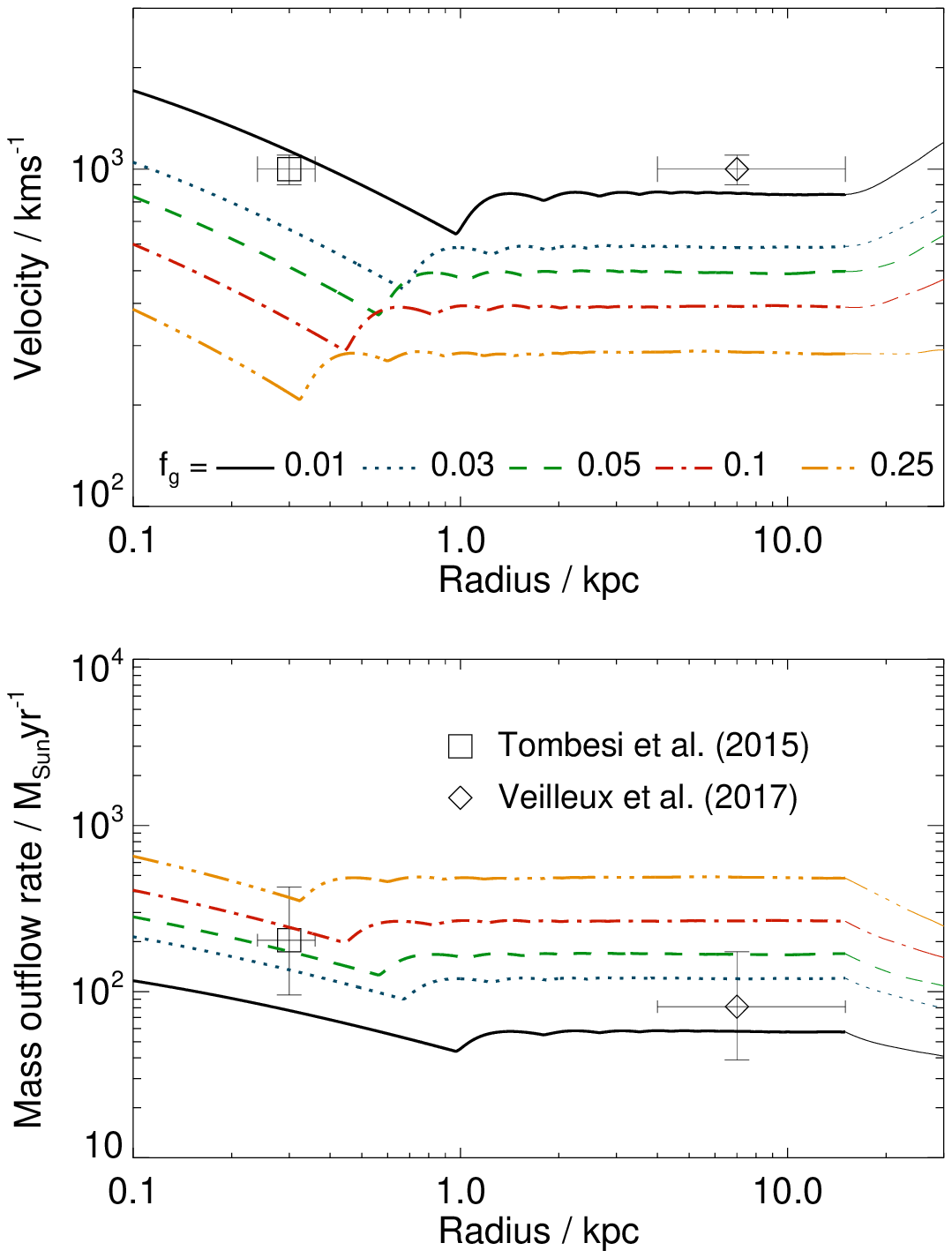}
    \caption{Same as Figure \ref{fig:F11119_largescale}, but for a varying AGN luminosity history with $\delta_{\rm AGN} = 0.42$.}
    \label{fig:F11119_rep_largescale}
\end{figure}

An AGN luminosity history consisting of multiple episodes might also explain the presence of spatially distinct outflow components in this galaxy. The extended outflow may have been inflated by an earlier series of AGN episodes, while the central one has been launched more recently, collecting the gas that had been too dense to be removed by the first outflow. Separate sub-components of the outflows may emerge due to different gas densities in different directions, which lead to some parts of the outflow moving faster than others. However, this is a very tentative interpretation; we consider it in more detail in the Discussion (Section \ref{sec:distinct_outflows}).

The situation in IRAS F11119+3257 is intermediate between the two previous cases. Figures \ref{fig:F11119_largescale} and \ref{fig:F11119_rep_largescale} show the expansion of the outflow in this galaxy for continuous and intermittent driving, respectively, with data points from \cite[squares]{Tombesi2015Natur} and \cite[diamonds]{Veilleux2017ApJ}. Once again, continuous driving is unable to reproduce both the velocity and mass outflow rate of the observed large-scale outflow, although the discrepancy is not as large as in the case of PDS 456. Meanwhile, the inner outflow at $R=300$~pc can be explained by the model, assuming that $0.05 < f_{\rm g} < 0.1$ and that the outflow does not cover the whole sky when looking from the AGN position. On the other hand, intermittent driving, with a high duty cycle $\delta_{\rm AGN} = 0.42$, produces reasonable agreement for the outer outflow, assuming a rather low gas fraction of $f_{\rm g} = 0.01$. This gas fraction corresponds to $M_{\rm g} = 10^9 \msun$ within $R = 15$~kpc, as observed \citep{Veilleux2017ApJ}. The inner outflow is much younger \citep[its approximate flow timescale is $t_{\rm fl} \sim 4\times10^5$~yr, while the outer outflow has $t_{\rm fl} \sim 7\times10^6$~yr, cf.][]{Veilleux2017ApJ} and may be driven by the current AGN episode, while the outer one is likely to be the product of numerous older episodes.

The difference between outflow parameters in continuous and intermittent AGN luminosity simulations is easy to understand qualitatively. Once the outflow expands beyond the central few hundred parsecs, its behaviour gradually begins to follow the time-averaged energy input from the AGN, rather than the instantaneous luminosity. If the AGN is strongly variable, the time-averaged input is significantly smaller than it would be if the AGN was radiating continuously at $L=L_{\rm max}$ or close to it. In fact, with the luminosity prescription we choose, the total energy emitted by the AGN from the beginning of an episode to the time its luminosity drops below $0.01 L_{\rm Edd}$ is
\begin{equation}
    E_{\rm out,r} \simeq 2.75 t_{\rm q} L_{\rm max} \simeq 0.057 t_{\rm r} \delta_{\rm AGN} L_{\rm max}.
\end{equation}
Note that we assumed $t_{\rm r} > t_{\rm ep}$ when calculating the above expression. A constant-luminosity AGN episode emits 
\begin{equation}
    E_{\rm out,c} = t_{\rm r} L_{\rm max}
\end{equation}
over a time $t_{\rm r}$. Therefore, an outflow driven by an intermittent AGN eventually behaves similarly to how it would if it were driven by a continuous energy injection at a fraction $0.057\delta_{\rm AGN}$ of the maximum rate. With the two values of duty cycle that we use, this corresponds to an energy input of $0.004$ and $0.024$ of the maximum rate. The outflow energy rate would then comprise the same fraction of the continuously-driven outflow energy rate, and the momentum rate would be a fraction $\left(E_{\rm out,r}/E_{\rm out,c}\right)^{2/3}$ of the continuously-driven one. If the current $L_{\rm AGN}$ is higher than its long-term average $0.057 \delta_{\rm AGN} L_{\rm max}$, the derived momentum and energy loading factors will be correspondingly lower than expected from analytical calculations. Conversely, if the AGN has recently faded and its current luminosity is much lower than the average, the observed outflow would appear abnormally powerful.

\section{Results: loading factor distribution} \label{sec:loading}

Individual object observations effectively represent only snapshots of the co-evolution of the AGN and its outflow. It is difficult to know which observations represent `typical' cases for outflows or even if such `typical' cases exist. However, by considering what values of momentum and energy loading factors would be observed at random times in our simulations, we can better understand how likely the observed low loading factors are.

\begin{figure}
	\includegraphics[width=\columnwidth]{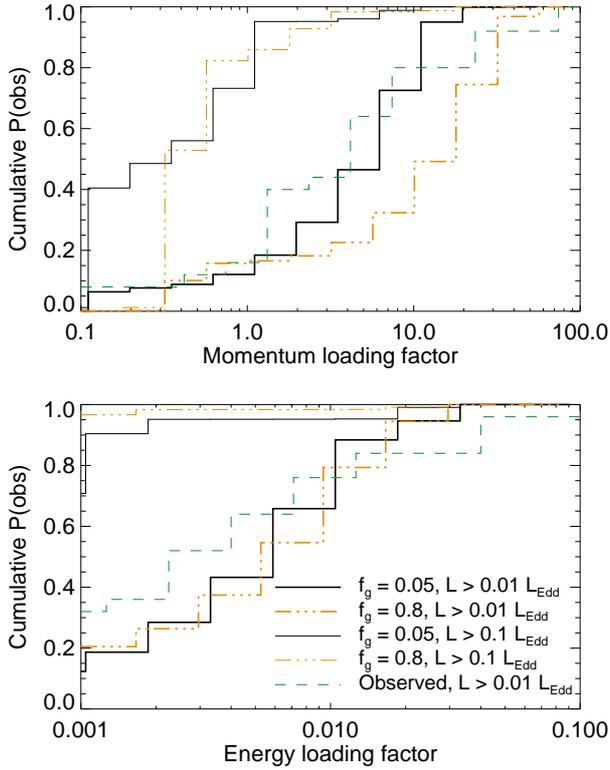}
    \caption{Cumulative distribution of observable instantenous momentum (top) and energy (bottom) loading factors in quasar (PDS 456) models with different gas fractions (black solid line for $f_{\rm g} = 0.05$, orange dash-triple-dotted line for $f_{\rm g} = 0.8$) and different AGN selection thresholds (thin lines for $L_{\rm AGN} > 0.1 L_{\rm Edd}$, thick lines for $L_{\rm AGN} > 0.01 L_{\rm Edd}$). This includes only outflows with $R_{\rm out} < 10$~kpc. Green dashed line shows loading factor distribution of observed molecular outflows in AGN with $L_{\rm AGN} > 0.01 L_{\rm Edd}$.}
    \label{fig:quasars_h}
\end{figure}

\begin{figure}
	\includegraphics[width=\columnwidth]{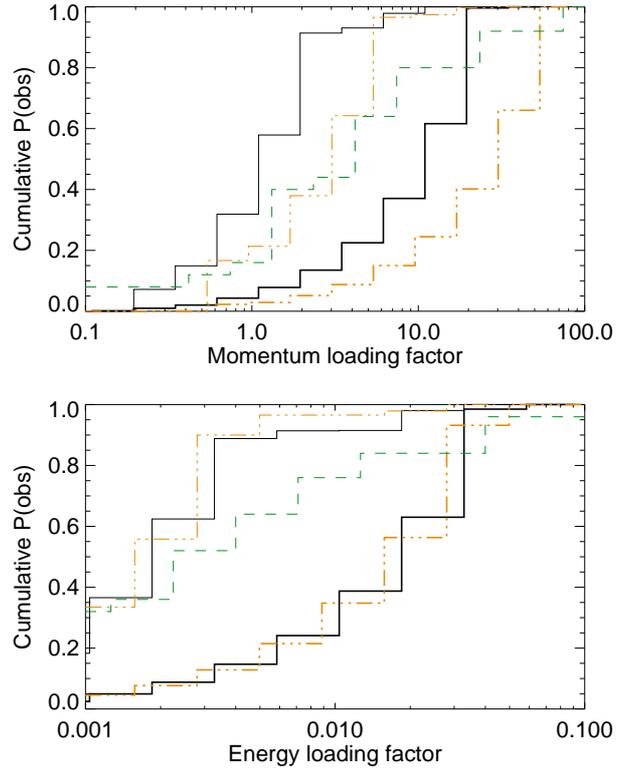}
    \caption{Same as Figure \ref{fig:quasars_h}, but with a five times longer AGN episode duration (five times higher duty cycle).}
    \label{fig:quasars_long_h}
\end{figure}

In order to produce a set of simulation results encompassing a greater variety of possible galaxy properties than those of Mrk 231, PDS 456 and IRAS F11119, we run simulations with two vastly different sets of galaxy properties. One set of properties is the same as those of PDS 456, while the other corresponds to a smaller galaxy with a much weaker Seyfert AGN (see the rightmost column in Table \ref{table:params}). Such an AGN has a 500 times lower luminosity than PDS 456 and produces outflows with velocities $100$~km~s$^{-1} < v_{\rm out} < 300$~km~s$^{-1}$ and mass flow rates $5 \, \msun$~yr$^{-1} < \dot{M}_{\rm out} < 40\, \msun$~yr$^{-1}$. However, the momentum and energy loading factor distributions are virtually indistinguishable between the Seyfert and PDS 456 models, provided that we choose the same threshold, in terms of Eddington ratio, for the selection of AGN (see below). Therefore, we only present the results of PDS 456 models. We also explore the effect of gas density by running simulations with gas fractions ranging from a relatively gas-poor $f_{\rm g} = 0.05$ to an extremely gas-rich $f_{\rm g} = 0.8$. We show the results of the two extreme cases; other cases produce loading factor distributions intermediate between these ones.

The loading factors that can be observed depend strongly on the threshold for detection and/or selection of AGN. Although realistic observational thresholds are based on luminosity (either bolometric or, more frequently, of a given indicator), physically, the Eddington ratio $f_{\rm Edd}$ is more important. Here, we present results using two selection thresholds: $L > 0.1 L_{\rm Edd}$ and $L > 0.01 L_{\rm Edd}$, where $L_{\rm Edd} = 1.7\times10^{47}$~erg~s$^{-1}$ for the PDS 456 models. The lower threshold corresponds approximately to the canonical switch of accretion mode from hot and spherical to a thin disc \citep{Best2012MNRAS, Sadowski2013MNRAS}. The higher threshold leads to only the brightest AGN being selected; this may represent a situation where only a sample of the very brightest AGN is selected by luminosity, since such galaxies must have both high SMBH masses and high $f_{\rm Edd}$. In addition, in all models, we only consider outflows within 10~kpc of the nucleus, as it is unlikely the AGN would be able to continue flickering on a $\sim 10^4$~yr timescale once most of the gas is removed well beyond the bulge \citep{Zubovas2016MNRASa}.

Figure \ref{fig:quasars_h} shows the expected cumulative distributions of momentum (top) and energy (bottom) loading factors that could be observed at random times. The different line colours correspond to models with the lowest ($f_{\rm g} = 0.05$, black solid) and highest ($f_{\rm g} = 0.8$, orange triple-dot-dashed) gas fractions. Line thicknesses correspond to different selection thresholds: $L > 0.01 L_{\rm Edd}$ (thick lines, lower probabilities) and $L > 0.1 L_{\rm Edd}$ (thin lines, higher probabilities). 

Looking at the AGN with $L > 0.01 L_{\rm Edd}$, we see that galaxies with higher gas fractions tend to have higher average momentum loading factors, therefore the probability of observing low loading factors is lower in these models; the energy loading factors show no dependence on gas fraction. It is very likely that a randomly-chosen system would show a momentum loading factor smaller than the analytical prediction $\dot{p} \simeq 20 L_{\rm AGN}/c$: in the lowest-density model, this probability is $P \sim 95\%$, and even in the highest-density model, it is still $P \sim 60\%$. The probability of detecting outflows with $\dot{p} < L_{\rm AGN}/c$ is small but non-negligible, $\sim 10-15\%$ depending on gas density. The instantaneous energy loading factors are almost uniformly distributed up to $E_{\rm load} \simeq 0.015$, also suggesting that observed values of $\dot{E} < 0.01 L_{\rm AGN}$ should be common ($P \sim 70\%$). The dearth of energy loading factors $>0.02$ is expected, because even with continuous driving, the kinetic energy of the outflow is expected to be only $\sim 1/3$ of the total energy transferred from the AGN \citep{Zubovas2012ApJ}, i.e. $\sim \eta/6 \sim 0.016$; the rest of the wind energy is used up to do work against gravity and $p$d$V$ work.

If we concentrate only on the brightest AGN, using a detection threshold $L > 0.1 L_{\rm Edd}$ (thin lines in the figure), the expected momentum- and energy loading factor distributions become much more extreme. The instantaneous momentum loading factor essentially never rises above $4$, even for the highest-density models, and the probability of observing $\dot{p} < L_{\rm AGN}/c$ is as high as $70-80\%$, almost independent of gas density. Energy loading factors are essentially always below $0.001$. The $\sim5\%$ of cases with higher energy loading factors correspond to the first AGN episode, when the outflow is very close to the nucleus.

For comparison, we plot the momentum and energy loading factor distribution of observed molecular outflows with $L_{\rm AGN} > 0.01 L_{\rm Edd}$ as green dashed histograms. The momentum loading factor distribution mostly follows the low-$f_{\rm g}$ model, as might be expected for local, generally gas-poor, galaxies. Some of the higher momentum-loading factors might be the result of higher gas fractions. The observed energy loading factor distribution is broader, with $\sim 20\%$ of observed outflows having $E_{\rm load} > 0.02$. As we show in the next subsection, this difference may be the result of real AGN having a variety of duty cycles.

\subsection{Effect of AGN duty cycle} \label{sec:duty}

In Figure \ref{fig:quasars_long_h}, we show the same result as in Figure \ref{fig:quasars_h}, but for models with a five times higher duty cycle $\delta_{\rm AGN} = 0.42$. In this case, all loading factor distributions are significantly broader, with the median momentum loading factor being $\dot{p} \sim 14 L_{\rm AGN}/c$ for $f_{\rm g} = 0.05$ and $\dot{p} \sim 30 L_{\rm AGN}/c$ for $f_{\rm g} = 0.8$. Meanwhile, energy loading factors tend to stay rather small, with the median $\dot{E}_{\rm out, med} \simeq 0.02 L_{\rm AGN}$, although occasionally the effective outflow energy rate rises as high as $\dot{E}_{\rm out} \sim 0.06 L_{\rm AGN}$. The probability of observing $\dot{p}_{\rm out} < L_{\rm AGN}/c$ is $2-6\%$. If only the brightest AGN are selected (thin lines), the typical loading factors decrease, in a similar fashion to the low-duty-cycle model, and the probability of observing $\dot{p}_{\rm out} < L_{\rm AGN}/c$ rises to $20-35\%$.

The differences between the loading factor distributions in the low- and high-duty-cycle models arise from the average outflow energy. The AGN energy input is 5 times higher in the high-$\delta_{\rm AGN}$ model, therefore the outflow energy (momentum) rate is also 5 ($5^{2/3} \sim 3$) times higher. The distribution of AGN luminosities, on the other hand, does not change between the two models, since the duty cycle only affects the duration of the quiescent period between episodes, which is not included when selecting the AGN for these mock observations. As a result, the average momentum (energy) loading factor that has the same cumulative probability is $\sim 3$ ($\sim 5$) times higher in the high-duty-cycle model than in the low-duty-cycle one, as seen in Figures \ref{fig:quasars_h} and \ref{fig:quasars_long_h}. During the quiescent phase, the outflow slows down more in the low-duty-cycle model, therefore once a new AGN episode begins, the instantaneous momentum and energy loading factors are much lower.

The distribution of observed loading factors (green dashed histograms) is broadly consistent with the models. The agreement is better when the loading factors are higher; conversely, the low-duty-cycle models fit the data better at the low-loading-factor end. This behaviour is qualitatively understandable: AGN with high duty cycles have more powerful outflows and therefore tend to show higher loading factors, so a mix of AGN with different duty cycles should show a distribution that matches both low- and high-duty cycle models at different parts of the distribution.

\section{Discussion} \label{sec:discuss}

\subsection{Correlation between outflows and UFOs}

Wide-angle disc winds (see Section \ref{sec:wind}) have been observed in many AGN; they are typically called ultra-fast outflows, or UFOs \citep{Tombesi2010AA, Tombesi2010ApJ, Tombesi2012MNRAS, Gofford2013MNRAS}. In Mrk 231 \citep{Feruglio2015AA}, PDS 456 \citep{Reeves2003ApJ,Bischetti2019AA} and IRAS F11119+3257 \citep{Tombesi2015Natur,Veilleux2017ApJ}, both wind and outflow have been observed simultaneously. In general, UFO properties correlate with those of the AGN much better than with the outflow. This is to be expected: the UFO is detected at distances $r_{\rm UFO} \ll 1$~pc from the nucleus and can react to changes in AGN luminosity on timescales of a few months or shorter \citep{Matzeu2017MNRAS}. In fact, most UFOs might have ages of several weeks or months \citep{King2015ARAA}.

The observed UFO kinetic power is very similar to that of the outflow in Mrk 231 \citep{Feruglio2015AA}, although this is not the case in IRAS F11119+3257 and PDS 456 \citep{Veilleux2017ApJ, Nardini2018MNRAS, Bischetti2019AA}. This situation echoes the correlations between outflow properties and AGN luminosity. Therefore, we feel confident in using AGN luminosity as an effective tracer of wind power transferred to the outflow.

In our models, we assume a constant $5\%$ efficiency of energy transfer from the AGN to the wind and, subsequently, to the outflow. Real UFOs show a spread of kinetic powers that range from $10^{-3} L_{\rm AGN}$ to $>0.1 L_{\rm AGN}$ \citep{Tombesi2013MNRAS, Fiore2017AA}. This spread may be the result of different SMBH spins, which lead to different radiative efficiencies \citep{King2015ARAA}, rapid UFO variability in time \citep{Pounds2016MNRAS, Reeves2018ApJa, Reeves2018ApJb}, clumps of different density moving into and out of the line of sight \citep{Matzeu2016MNRAS,Reeves2020arXiv}, different viewing angles \citep{Matthews2016MNRAS,Giustini2019AA} or other differences in the details of wind geometries. Launching mechanisms are very important as well: at high Eddington ratios, continuum driving might be powerful enough to drive the wind \citep{King2010MNRASb}, while at lower luminosities, processes such as line driving and magnetic driving become progressively more important \citep{Proga2000ApJ, Proga2003ApJ, Sim2008MNRAS, Fukumura2010ApJ, Higginbottom2014ApJ, Nomura2016PASJ, Cui2020ApJ}. In these cases, the force multiplier term (the ratio of driving force to the AGN radiation pressure force) depends on several other factors that may differ from one system to another.

Knowledge of outflow properties might provide a way to constrain the origin and physics of UFOs in the near future, when the number of objects with simultaneous UFO and large-scale outflow detections increases. If the spread of UFO kinetic powers is due to variability on short timescales of decades or less, or a result of different viewing directions, we should also see a large scatter in the correlations between UFO and outflow properties, since outflows react to the long-term energy input by the UFO. Conversely, if UFO differences arise from a diversity of launching mechanisms or efficiencies, then outflow properties might show stronger and less scattered correlations with UFO properties than with AGN luminosity or Eddington ratio.

\subsection{Outflow correlations with other galaxy parameters} \label{sec:galaxy_correl}

The variation of AGN luminosity on $10^3-10^6$~yr timescales induces significant scatter in the correlations between outflow parameters and $L_{\rm AGN}$, as we have shown. However, analytical models \citep[cf.][]{Zubovas2012ApJ} predict other correlations between outflow properties and host galaxy parameters, or between separate outflow parameters, that should be preserved and help constrain whether outflows are actually energy-driven. 

For example, since the energy transferred to the outflow is independent of gas density, we expect outflows in denser systems to be slower, with $v_{\rm out} \propto f_{\rm g}^{-1/3} \propto M_{\rm gas}^{-1/3}$. This relation is as strong as the dependence on the time-averaged AGN luminosity $\langle L_{\rm AGN}\rangle$: $v_{\rm out} \propto \langle L_{\rm AGN}\rangle^{1/3}$. Therefore, we expect the correlation with $M_{\rm gas}$ to be noticeable even accounting for the scatter induced by AGN luminosity variations. 

The mass outflow rate $\dot{M}_{\rm out} \propto f_{\rm g} v_{\rm out} \propto f_{\rm g}^{2/3} \propto M_{\rm gas}^{2/3}$ should have an even stronger correlation with gas mass than outflow velocity, therefore the correlation should be even more evident. Of course, the multiphase nature of the gas introduces more scatter into the correlations: since not all the outflowing gas might be detected, and not all the detected gas might be susceptible to AGN wind driving, the relation between the two quantities may be rather loose. Future observations will improve our understanding of these correlations in two ways: by identifying diffuse gas mass and/or its density and detecting multiple outflow phases; and by determining the radial variation of outflow properties, in a similar way to what has been attempted for Mrk 231 \citep[and Section \ref{sec:largescale}]{Feruglio2015AA}. Although observationally demanding, the full capabilities of {\it ALMA} have not yet been exploited in the latter sense. The spatial resolution achieved in the {\it ALMA} observations of PDS 456 and IRAS F11119, for instance, is $\sim$0.7 and 2.8 kpc, respectively, whereas resolutions of 0.1 kpc or slightly better can be, in principle, reached. Radial profiles of ionized outflows, which can now be obtained for a few nearby Seyferts only (e.g. \citealt{Crenshaw2015ApJ,Venturi2018AA}), will be routinely reconstructed in a few years time through integral field spectroscopy with the {\it Extremely Large Telescope}. On the other hand, the {\it Square Kilometre Array} could eventually reveal the direct radio emission from the shocked wind and ambient gas.

The correlations presented above are technically only valid if the gas and background potential are distributed isothermally. Different radial profiles would affect the correlations, but the qualitative trends should remain the same. It is worth noting that mass outflow rate and outflow velocity should correlate with each other independently of radial density profile or AGN luminosity, and the correlation should only depend on gas mass in the galaxy. Utilising this correlation might help determine the fraction of total outflowing material that is detected directly and/or the fraction of all gas that is susceptible to joining the outflow.

Finally, there may be a systematic difference between outflows detected close to the nucleus and those seen further away. A single AGN episode may inflate an outflow to a radius $R_1 \sim v_{\rm out} t_{\rm ep} \sim 1 v_{1000} t_{\rm Myr}$~kpc, where $t_{\rm Myr}$ is AGN episode duration in Myr. Given that outflow velocities are usually below $1000$~km~s$^{-1}$ and durations of individual episodes should not reach 1 Myr, only outflows detected within a few hundred parsecs of the nucleus might be expected to be inflated by the current AGN episode. These outflows might generally be faster for a given galaxy gas content, carry more mass and show a stronger correlation with the current AGN luminosity than outflows detected further out, where multiple AGN episodes had contributed to their present-day properties. The two outflows in IRAS F11119+3257 reflect these differences rather well, with the inner outflow more easily explained by a model with continuous AGN driving and the outer one requiring intermittent driving \citep[Section \ref{sec:largescale}; see also][]{Nardini2018MNRAS}). The properties of outflows distant from the nucleus might correlate better with the mass of the SMBH, which sets the maximum, and hence probably the average, AGN luminosity, as found by \cite{Gonzalez2017ApJ}; we explore this connection in a companion paper (Nardini \& Zubovas, in prep.). We predict that as outflow data becomes better constrained, differences between close-in and distant massive outflows should emerge.

\subsection{Constraining AGN luminosity histories and duty cycles} \label{sec:duty_cycle_estimates}

Our results may be useful in inferring the luminosity histories and duty cycles of AGN over the lifetime of an observed outflow, which may be a few times $10^5-10^7$~yr. In order to constrain the luminosity history, we can compare the predicted distribution of momentum and/or energy loading factors with the observed distribution in a suitably selected subsample. Our models show that a value of $f_{\rm Edd}$ is a better criterion for selecting the AGN, rather than $L_{\rm AGN}$. While using the Eddington ratio instead of absolute luminosity requires knowledge of the mass of the SMBH powering the outflow, thus reducing the available observational sample and introducing additional uncertainty, this choice allows us to investigate loading factor distributions in a physically meaningful way.

The predicted momentum and energy loading factor distributions (Figures \ref{fig:quasars_h} and \ref{fig:quasars_long_h}) show that every value, up to some maximum, is approximately equally likely to be observed. The maximum value depends on the adopted $f_{\rm Edd}$ threshold and the duty cycle, with lower thresholds and higher duty cycles leading to higher maximum loading factor values. In addition, momentum loading, but not energy loading, depends on host galaxy gas density, with the maximum value $\propto f_{\rm g}^{1/3}$, as expected from analytical estimates (see Section \ref{sec:galaxy_correl}). The flatness of these distributions relies on the fact that AGN luminosity decreases approximately as a power law during each episode. If the AGN decayed faster, e.g. exponentially, we would expect many more high loading factors, while if the luminosity stayed at an approximately constant high level for a significant fraction of each episode, we would expect more low loading factors \citep{Zubovas2018MNRAS}. In principle, then, knowing the distribution of loading factors in a real outflow sample, we can constrain the general shape of the AGN luminosity variation with time.

Real outflows have momentum loading factors that range from well below unity to $\sim 10^3$ (see Table \ref{ts}), with some fossil outflows having values as large as $>10^4$ \citep{Fluetsch2019MNRAS}, however most are clustered in the range $1 < p_{\rm load} < 30$. In this range, the values are distributed mostly uniformly. This allows us to tentatively suggest that most galaxies with known outflows have high gas fractions and/or rather high AGN duty cycles over the outflow lifetime. We refrain from attempting any quantitative estimates because of the rather small sample size.

Assuming that the adopted AGN luminosity prescription is qualitatively correct, we can put constraints on the AGN duty cycle using an individual outflow observation. The average AGN luminosity over a single episode is $L_{\rm ave} \simeq 2.75 L_{\rm max} t_{\rm q}/t_{\rm ep} \simeq 0.057 L_{\rm max}$, where the maximum luminosity is probably $\sim L_{\rm Edd}$. The long-term average AGN luminosity is then $\langle L_{\rm AGN} \rangle \simeq \delta_{\rm AGN} L_{\rm ave} \simeq 0.057 \delta_{\rm AGN} L_{\rm Edd}$. Given that the kinetic energy of the outflow is $\dot{E}_{\rm out} \simeq 0.02 \langle L_{\rm AGN}\rangle$ \citep{Zubovas2012ApJ}, we can estimate $\delta_{\rm AGN}$:
\begin{equation}
    \dot{E}_{\rm out} \simeq 0.02 \langle L_{\rm AGN}\rangle \sim 10^{-3} \delta_{\rm AGN} L_{\rm Edd};
\end{equation}
\begin{equation} \label{eq:delta}
    \delta_{\rm AGN} \sim 10^3 \frac{\dot{E}_{\rm out}}{L_{\rm Edd}}.
\end{equation}

This relation is not appropriate for outflows very close to the nucleus, where the outflow properties are dependent on the most recent episode rather than the long-term average (see Section \ref{sec:galaxy_correl}). For outflows more than a few hundred parsecs away from the nucleus, however, this may be applicable. For these outflows, a high kinetic energy value $\dot{E}_{\rm out}/ L_{\rm Edd} > 10^{-3}$ indicates that the AGN was shining essentially continuously while the outflow was expanding, while a lower value indicates a more sporadic AGN history. The individual galaxies we investigated, Mrk 231, PDS 456 and IRAS F11119+3257, have, respectively, $\dot{E}_{\rm out}/L_{\rm Edd} \simeq 6\times10^{-3}$, $7\times 10^{-5}$ and $2\times10^{-3}$ (for the outer outflow in the latter two cases). Mrk 231 is clearly consistent with continuous driving by an AGN, with formal $\delta_{\rm AGN} > 5$. On the other hand, PDS 456 requires $\delta_{\rm AGN} \sim 0.07$, very close to the one we used in our model. IRAS F11119+3257 has $\delta_{\rm AGN} = 1.85$, higher than our adopted $0.42$, although we also used a higher $L_{\rm max}$ in this case.

It is important to keep in mind that this duty cycle estimate is only applicable for the period during which the outflow has been inflated. An SMBH might grow in `spurts', periods lasting several tens of Myr \citep{Yu2002MNRAS, Hopkins2005ApJ}, each composed of many shorter `flickering' episodes lasting $\sim 10^5$~yr each. The duty cycle estimate from eq. \ref{eq:delta} only relates to the period within such a `spurt'. Over a much longer timescale comparable to the Hubble time, the duty cycle will be significantly smaller, consistent with observational and theoretical estimates \citep{Wang2006ApJ,Shankar2013MNRAS}.

\subsection{Multiple spatially distinct outflows} \label{sec:distinct_outflows}

Our model explicitly assumes that the outflow is singular, i.e. there are no spatially distinct components. Physically, this assumption means that the outflow expands through single-phase diffuse ISM without affecting dense clouds, and that those clouds remain dense after the outflow passage and do not refill the cavity left behind. Evidently, this is not always the case: for example, PDS 456 has both a central and an extended outflow component \citep{Bischetti2019AA}, while IRAS F11119+3257 shows spatially very distinct outflow signatures in OH and CO data \citep{Tombesi2015Natur,Veilleux2017ApJ}. Although alternative interpretations of the data are possible, the plausible existence of separate outflows on different scales must be considered.

Nevertheless, we are confident such a situation should be rare, and most outflows should have only a single spatial component, for the following reason. Consider a galaxy spheroid with a diffuse gas fraction $f_{\rm g,0}$ and some number of dense clouds. An AGN episode drives a large-scale outflow through the diffuse medium, leaving a cavity behind. The cavity is filled with the shocked AGN wind and has $f_{\rm g, cav} \ll f_{\rm g,0}$, i.e. it is effectively empty, but overpressurised with respect to the initial ISM pressure. While the AGN episode continues, this shocked wind pressure effectively confines the dense clouds, potentially disrupting them and pushing them away with the diffuse outflow \citep{Hopkins2010MNRAS} or compressing them leading to enhanced star formation \citep{Zubovas2014MNRASc}. Once the episode ends, the wind cools down; if the shocked wind is almost adiabatic, cooling happens on the outflow dynamical timescale $t_{\rm d} \simeq R_{\rm out}/v_{\rm out} \simeq 10^6 R_{\rm kpc} v_{1000}^{-1}$~yr. At some point, dense clouds are no longer compressed and may evaporate on a timescale
\begin{equation}
    t_{\rm evap} \simeq 5\times10^5 R_{\rm pc} T_7^{-5/2} \, {\rm yr},
\end{equation}
where $T_7 \equiv T_{\rm sh}/(10^7 \,{\rm K})$ is the shocked wind temperature. Here we used the classical evaporation rate for a cloud in hot gas \citep{Cowie1977ApJ} and the \cite{Larson1981MNRAS} relations to connect typical cloud density and radius. This is likely a lower limit for the timescale: most clouds are larger, the wind temperature may well be lower once the clouds are no longer confined, and cloud compression during the AGN episode may lead to them having higher densities than those obtained from Larson relations. Therefore, clouds evaporate rather slowly and increase the diffuse gas density in the outflow cavity to some value $f_{\rm g,1} < f_{\rm g,0}$. Even if clouds evaporate rapidly, pressure equilibrium is reached once gas density in the cavity rises to the initial value $f_{\rm g,0}$, so $f_{\rm g,1} \leq f_{\rm g,0}$ always.

Once a new AGN episode begins, the wind encounters gas with, most likely, much lower density than the initial one. The resulting outflow is therefore faster than the (coasting) original outflow, and catches up with the latter on a timescale shorter than $t_{\rm d}$. If several episodes happen in quick succession, the cavity does not refill significantly, and all episodes drive the same outflow. On the other hand, if there is a very long gap between episodes, the fossil outflow may slow down, disperse and become no longer detectable. This occurs if the AGN is inactive for a period $t_{\rm off} \simgt 10 t_{\rm ep}$ \citep[cf.][]{King2011MNRAS}, i.e. if $\delta_{\rm AGN} < 0.1$ over the several-Myr timescale. Our estimate of the duty cycle in PDS 456 falls below this threshold, but not significantly so, therefore this galaxy may be one of the rare cases where the cavity evacuated by the first outflow has been refilled significantly by the time the second outflow started.

If the AGN is observed during the time when the inner outflow has not yet caught up with the outer one, the two outflows should show systematic differences. The inner outflow should be faster but carry less mass than the outer one. It may, however, be more difficult to detect all the mass contained in the outer stalling outflow, especially if its velocity is currently similar to the velocity dispersion and/or rotational velocity of other gas in the galaxy.

The above picture is complicated by the presence of coherent structures feeding the SMBH via gravitational torques \citep[e.g.,][]{Angles2017MNRAS}. They may continuously refill the outflow cavity faster than evaporating clouds do, leading to separate AGN episodes acting on distinct gas reservoirs and producing separate outflows. This situation is particularly relevant for gas-rich mergers, which are common hosts of AGN-driven outflows.

\subsection{Multiphase outflows} \label{sec:multiphase}

The outflowing gas can be ionised by shocks and by the AGN radiation field, and cools down mainly via bremsstrahlung, metal line and atomic cooling processes. The effect of shocks depends primarily on the outflow velocity and so remains almost constant as the outflow evolves. However, the much more abrupt changes in AGN luminosity can have significant effects on the ionisation state of the outflow. Even in the presence of the AGN radiation field, the outflow can cool and form molecules \citep{Zubovas2014MNRASa, Richings2018MNRAS, Richings2018MNRASb}. Higher AGN luminosity results in slower cooling, with $L_{\rm AGN} = 10^{45}$~erg~s$^{-1}$ leading to molecule formation on a timescale of $t_{\rm form} \simeq 0.3$~Myr, while for $L_{\rm AGN} = 10^{46}$~erg~s$^{-1}$, $t_{\rm form} \simeq 0.6$~Myr, at least in models with constant ISM density $n_{\rm H} = 10$~cm$^{-3}$ \citep{Richings2018MNRAS,Richings2018MNRASb}. These timescales are longer than the duration of individual AGN episodes estimated via observational \citep{Schawinski2015MNRAS} and analytical \citep{King2015MNRAS} arguments, therefore we may expect the ratio of molecular and atomic/ionised gas in outflows in real galaxies to be the result of a complex interplay of time-dependent heating and cooling processes. Evidently, at higher luminosity, the importance of heating increases: in AGN with $L_{\rm AGN} = 10^{45}$~erg~s$^{-1}$, ionised mass outflow rates are $\sim 0.01$ times the molecular ones, but the ratio increases to $>0.1$ for AGN with $L_{\rm AGN} = 10^{46}$~erg~s$^{-1}$ \citep{Fiore2017AA, Bischetti2019AA}. 

Qualitatively, this trend is easy to understand. When the AGN switches on, the heating rate increases significantly on a light-travel time. Meanwhile, the cooling rate hardly changes, since outflow properties change very little over the course of a single episode, assuming that the outflow has been expanding for a few episodes already. As the AGN luminosity decays and especially when the episode ends, the ionisation balance changes and may lead to the ionised gas disappearing completely. The recombination time in the comparatively dense outflow is much shorter than in the undisturbed ISM or circumgalactic medium (CGM) \citep{Zubovas2014MNRASa}, so we would not expect long-lived ionised remnants to be present in the outflow. Formation of molecules from the recently-cooled material takes some time, so fossil outflows might trace only the molecular component of the original outflow. This situation may result in ionised outflows having stronger correlations with $L_{\rm AGN}$ than molecular ones, simply because their {\em detection} is affected by the AGN more strongly than that of molecular gas. We plan to explore these correlations in more detail in a future publication.

\subsection{Implementation in numerical simulations}

Numerical simulations of galaxy evolution often include AGN feedback in the form of energy injection into gas surrounding the SMBH particle \citep[e.g.][]{Sijacki2007MNRAS, Vogelsberger2014MNRAS, Schaye2015MNRAS, Tremmel2019MNRAS}. Currently, these simulations reach spatial and temporal resolution that is comparable to the accreted mass and duration of individual AGN episodes \citep{Grand2017MNRAS, Tremmel2019MNRAS, Nelson2019MNRAS}. This presents an opportunity to upgrade the AGN feedback prescription in order to achieve more realistic galaxy behaviour. Two ingredients are necessary for this upgrade.

First of all, SMBH feeding is typically unresolved, with a Bondi-Hoyle-like prescription used to estimate the accretion rate \citep{Booth2009MNRAS}. As the mass, if not spatial, resolution reaches values low enough to accommodate the growth of an accretion disc around the SMBH (i.e. $m_{\rm res} \ll M_{\rm BH}$), introducing a sub-resolution prescription for disc evolution should give more realistic time evolution of the SMBH feeding rate and hence the variation of AGN luminosity. In particular, the accretion disc particle method \citep{Power2011MNRASb} can reproduce realistic AGN behaviour which has little dependence on the free parameters of the model \citep{Wurster2013MNRAS}. This prescription should be able to reproduce the AGN flickering behaviour on $10^4-10^5$~yr timescales \citep{King2015MNRAS, Schawinski2015MNRAS}, which in turn produces realistic distributions of outflow loading factors \citep[see also][]{Zubovas2018MNRAS}.

The second improvement would be to adopt a more realistic geometry of feedback energy injection. Depositing the energy into some number of nearest particles to the SMBH results in dense gas close to the SMBH absorbing a disproportionately large amount of energy, leading to weak outflows and unphysical stalling of SMBH growth \citep{Zubovas2016MNRASa}; this can be alleviated if energy is injected into a bicone. Using such a prescription would create more realistic dynamics of AGN feeding reservoirs and allow prolonged AGN growth periods that might contain numerous flickering episodes.

This approach would help make simulations more realistic on the level of individual galaxies and processes happening therein. Such simulations could be checked and constrained by the growing data set of observed AGN-driven outflows. Furthermore, they would help investigate the relationship between outflow parameters and the properties of their host galaxies and/or broader environment.

\section{Summary and conclusion} \label{sec:summary}

We numerically investigated the evolution of galactic outflows driven by intermittent AGN episodes each lasting $\sim 10^4-10^5$~yr, with particular emphasis on how this luminosity evolution affects the observable momentum and energy loading factors. We considered three individual galaxies - Mrk 231, PDS 456 and IRAS F11119+3257 (Section \ref{sec:largescale}) - and a distribution of loading factors seen in several tens of other outflows (Section \ref{sec:loading}). The main results are the following:

\begin{itemize}
    \item The outflow in Mrk 231, which shows a momentum loading factor of $\dot{p}_{\rm out}c/L_{\rm AGN} > 5$ and an energy loading factor $\dot{E}_{\rm out}/L_{\rm AGN} > 0.006$ (for the molecular component; both loading factors are presumably a factor $\sim 2$ greater when the neutral gas outflow is taken into account), can be adequately explained as an almost spherically symmetric energy-driven outflow expanding under continuous driving by the AGN for the past $t_{\rm out} \sim 1$~Myr.
    \item Conversely, the `unexpectedly weak' outflow in PDS 456 is not compatible with continuous driving at the present-day AGN luminosity, but can be explained using a model where the AGN luminosity varies in time with a duty cycle $\delta_{\rm AGN} = 0.084$; in this case, the present-day luminosity is much higher than the long-term average.
    \item The outflow in IRAS F11119+3257 is an intermediate case: the $300$-pc-scale outflow can be explained by a continuous-luminosity model, while the intermittent AGN model, with a duty cycle $\delta_{\rm AGN} = 0.42$, is a better fit to the $7$-kpc-scale one.
    \item For a population of AGN outflows observed at random times in their evolution, the distribution of observed momentum and energy loading factors does not depend on SMBH or galaxy mass, but only on the AGN duty cycle, gas density and the threshold $f_{\rm Edd}$ for AGN selection;
    \item Assuming that AGN are selected because of bright thin-disc emission, i.e. $f_{\rm Edd} > 0.01$, a small, but non-negligible fraction of outflows should have low momentum and energy loading factors: $p_{\rm load} < 1$ in $5-15\%$ of cases in models with low AGN duty cycle $\delta_{\rm AGN} = 0.084$, depending on gas density, with higher density leading to lower probability of observing small loading factors.
    \item The fraction of outflows with small loading factors increases significantly if only AGN with $f_{\rm Edd} > 0.1$ are considered ($p_{\rm load} < 1$ in $60-95\%$ of cases) and decreases if the AGN duty cycle is higher ($p_{\rm load} < 1$ in $2-6\%$ of cases when $\delta_{\rm AGN} = 0.42$).
\end{itemize}

These results show that a lot of the observed variation among AGN-driven outflows may simply be the outcome of AGN luminosity varying significantly with time. As more observational data is collected and as numerical simulation resolution steadily improves, the evolution of individual AGN outflows, as well as AGN luminosity histories over the past several Myr, should become traceable, greatly enhancing our understanding of the coevolution of SMBHs and their host galaxies.

\section*{Acknowledgements}

We thank Andrew King for insightful comments on the draft version of the paper. KZ is funded by the Research Council Lithuania grant no. S-MIP-20-43. EN acknowledges financial contribution from the agreement ASI-INAF n.2017-14-H.0 and partial support from the EU Horizon 2020 Marie Sk\l{}odowska-Curie grant agreement no. 664931.

\section*{Data availability}

Observational data used in this paper is presented in Section \ref{sec:data} and Table \ref{ts}; it is taken from publicly available sources. The code used for the modelling is currently being prepared for public release; for the moment, it is available upon reasonable request to the corresponding author.





\bsp	
\label{lastpage}

\begin{thebibliography}{}
\makeatletter
\relax
\def\mn@urlcharsother{\let\do\@makeother \do\$\do\&\do\#\do\^\do\_\do\%\do\~}
\def\mn@doi{\begingroup\mn@urlcharsother \@ifnextchar [ {\mn@doi@}
  {\mn@doi@[]}}
\def\mn@doi@[#1]#2{\def\@tempa{#1}\ifx\@tempa\@empty \href
  {http://dx.doi.org/#2} {doi:#2}\else \href {http://dx.doi.org/#2} {#1}\fi
  \endgroup}
\def\mn@eprint#1#2{\mn@eprint@#1:#2::\@nil}
\def\mn@eprint@arXiv#1{\href {http://arxiv.org/abs/#1} {{\tt arXiv:#1}}}
\def\mn@eprint@dblp#1{\href {http://dblp.uni-trier.de/rec/bibtex/#1.xml}
  {dblp:#1}}
\def\mn@eprint@#1:#2:#3:#4\@nil{\def\@tempa {#1}\def\@tempb {#2}\def\@tempc
  {#3}\ifx \@tempc \@empty \let \@tempc \@tempb \let \@tempb \@tempa \fi \ifx
  \@tempb \@empty \def\@tempb {arXiv}\fi \@ifundefined
  {mn@eprint@\@tempb}{\@tempb:\@tempc}{\expandafter \expandafter \csname
  mn@eprint@\@tempb\endcsname \expandafter{\@tempc}}}

\bibitem[\protect\citeauthoryear{{Aalto}, {Muller}, {Sakamoto}, {Gallagher},
  {Mart{\'\i}n}  \& {Costagliola}}{{Aalto} et~al.}{2012}]{Aalto2012AA}
{Aalto} S.,  {Muller} S.,  {Sakamoto} K.,  {Gallagher} J.~S.,  {Mart{\'\i}n}
  S.,   {Costagliola} F.,  2012, \mn@doi [\aap] {10.1051/0004-6361/201118052},
  \href {https://ui.adsabs.harvard.edu/abs/2012A&A...546A..68A} {546, A68}

\bibitem[\protect\citeauthoryear{{Angl{\'e}s-Alc{\'a}zar}, {Dav{\'e}},
  {Faucher-Gigu{\`e}re}, {{\"O}zel}  \& {Hopkins}}{{Angl{\'e}s-Alc{\'a}zar}
  et~al.}{2017}]{Angles2017MNRAS}
{Angl{\'e}s-Alc{\'a}zar} D.,  {Dav{\'e}} R.,  {Faucher-Gigu{\`e}re} C.-A.,
  {{\"O}zel} F.,   {Hopkins} P.~F.,  2017, \mn@doi [\mnras]
  {10.1093/mnras/stw2565}, \href
  {http://adsabs.harvard.edu/abs/2017MNRAS.464.2840A} {464, 2840}

\bibitem[\protect\citeauthoryear{{Barcos-Mu{\~n}oz} et~al.,}{{Barcos-Mu{\~n}oz}
  et~al.}{2018}]{Barcos2018ApJ}
{Barcos-Mu{\~n}oz} L.,  et~al., 2018, \mn@doi [\apjl]
  {10.3847/2041-8213/aaa28d}, \href
  {https://ui.adsabs.harvard.edu/abs/2018ApJ...853L..28B} {853, L28}

\bibitem[\protect\citeauthoryear{{Bentz} et~al.,}{{Bentz}
  et~al.}{2013}]{Bentz2013ApJ}
{Bentz} M.~C.,  et~al., 2013, \mn@doi [\apj] {10.1088/0004-637X/767/2/149},
  \href {https://ui.adsabs.harvard.edu/abs/2013ApJ...767..149B} {767, 149}

\bibitem[\protect\citeauthoryear{{Best} \& {Heckman}}{{Best} \&
  {Heckman}}{2012}]{Best2012MNRAS}
{Best} P.~N.,  {Heckman} T.~M.,  2012, \mn@doi [\mnras]
  {10.1111/j.1365-2966.2012.20414.x}, \href
  {https://ui.adsabs.harvard.edu/abs/2012MNRAS.421.1569B} {421, 1569}

\bibitem[\protect\citeauthoryear{{Bischetti} et~al.,}{{Bischetti}
  et~al.}{2019}]{Bischetti2019AA}
{Bischetti} M.,  et~al., 2019, \mn@doi [\aap] {10.1051/0004-6361/201935524},
  \href {https://ui.adsabs.harvard.edu/abs/2019AA...628A.118B} {628, A118}

\bibitem[\protect\citeauthoryear{{Bolatto}, {Wolfire}  \& {Leroy}}{{Bolatto}
  et~al.}{2013a}]{Bolatto2013ARAA}
{Bolatto} A.~D.,  {Wolfire} M.,   {Leroy} A.~K.,  2013a, \mn@doi [\araa]
  {10.1146/annurev-astro-082812-140944}, \href
  {https://ui.adsabs.harvard.edu/abs/2013ARAA..51..207B} {51, 207}

\bibitem[\protect\citeauthoryear{{Bolatto} et~al.,}{{Bolatto}
  et~al.}{2013b}]{Bolatto2013Nat}
{Bolatto} A.~D.,  et~al., 2013b, \mn@doi [\nat] {10.1038/nature12351}, \href
  {https://ui.adsabs.harvard.edu/abs/2013Natur.499..450B} {499, 450}

\bibitem[\protect\citeauthoryear{{Booth} \& {Schaye}}{{Booth} \&
  {Schaye}}{2009}]{Booth2009MNRAS}
{Booth} C.~M.,  {Schaye} J.,  2009, \mn@doi [\mnras]
  {10.1111/j.1365-2966.2009.15043.x}, \href
  {http://adsabs.harvard.edu/abs/2009MNRAS.398...53B} {398, 53}

\bibitem[\protect\citeauthoryear{{Bourne} \& {Nayakshin}}{{Bourne} \&
  {Nayakshin}}{2013}]{Bourne2013MNRAS}
{Bourne} M.~A.,  {Nayakshin} S.,  2013, \mn@doi [\mnras]
  {10.1093/mnras/stt1739}, \href
  {http://adsabs.harvard.edu/abs/2013MNRAS.436.2346B} {436, 2346}

\bibitem[\protect\citeauthoryear{{Bower}, {Benson}, {Malbon}, {Helly}, {Frenk},
  {Baugh}, {Cole}  \& {Lacey}}{{Bower} et~al.}{2006}]{Bower2006MNRAS}
{Bower} R.~G.,  {Benson} A.~J.,  {Malbon} R.,  {Helly} J.~C.,  {Frenk} C.~S.,
  {Baugh} C.~M.,  {Cole} S.,   {Lacey} C.~G.,  2006, \mn@doi [\mnras]
  {10.1111/j.1365-2966.2006.10519.x}, \href
  {http://adsabs.harvard.edu/abs/2006MNRAS.370..645B} {370, 645}

\bibitem[\protect\citeauthoryear{{Cicone}, {Feruglio}, {Maiolino}, {Fiore},
  {Piconcelli}, {Menci}, {Aussel}  \& {Sturm}}{{Cicone}
  et~al.}{2012}]{Cicone2012AA}
{Cicone} C.,  {Feruglio} C.,  {Maiolino} R.,  {Fiore} F.,  {Piconcelli} E.,
  {Menci} N.,  {Aussel} H.,   {Sturm} E.,  2012, \mn@doi [\aap]
  {10.1051/0004-6361/201218793}, \href
  {http://adsabs.harvard.edu/abs/2012A%26A...543A..99C} {543, A99}

\bibitem[\protect\citeauthoryear{{Cicone} et~al.,}{{Cicone}
  et~al.}{2014}]{Cicone2014AA}
{Cicone} C.,  et~al., 2014, \mn@doi [\aap] {10.1051/0004-6361/201322464}, \href
  {http://adsabs.harvard.edu/abs/2014A%26A...562A..21C} {562, A21}

\bibitem[\protect\citeauthoryear{{Cicone} et~al.,}{{Cicone}
  et~al.}{2018}]{Cicone2018ApJ}
{Cicone} C.,  et~al., 2018, \mn@doi [\apj] {10.3847/1538-4357/aad32a}, \href
  {https://ui.adsabs.harvard.edu/abs/2018ApJ...863..143C} {863, 143}

\bibitem[\protect\citeauthoryear{{Combes} et~al.,}{{Combes}
  et~al.}{2013}]{Combes2013AA}
{Combes} F.,  et~al., 2013, \mn@doi [\aap] {10.1051/0004-6361/201322288}, \href
  {https://ui.adsabs.harvard.edu/abs/2013A&A...558A.124C} {558, A124}

\bibitem[\protect\citeauthoryear{{Costa}, {Rosdahl}, {Sijacki}  \&
  {Haehnelt}}{{Costa} et~al.}{2018}]{Costa2018MNRAS}
{Costa} T.,  {Rosdahl} J.,  {Sijacki} D.,   {Haehnelt} M.~G.,  2018, \mn@doi
  [\mnras] {10.1093/mnras/stx2598}, \href
  {http://adsabs.harvard.edu/abs/2018MNRAS.473.4197C} {473, 4197}

\bibitem[\protect\citeauthoryear{{Cowie} \& {McKee}}{{Cowie} \&
  {McKee}}{1977}]{Cowie1977ApJ}
{Cowie} L.~L.,  {McKee} C.~F.,  1977, \mn@doi [\apj] {10.1086/154911}, \href
  {http://adsabs.harvard.edu/abs/1977ApJ...211..135C} {211, 135}

\bibitem[\protect\citeauthoryear{{Crenshaw}, {Fischer}, {Kraemer}  \&
  {Schmitt}}{{Crenshaw} et~al.}{2015}]{Crenshaw2015ApJ}
{Crenshaw} D.~M.,  {Fischer} T.~C.,  {Kraemer} S.~B.,   {Schmitt} H.~R.,  2015,
  \mn@doi [\apj] {10.1088/0004-637X/799/1/83}, \href
  {https://ui.adsabs.harvard.edu/abs/2015ApJ...799...83C} {799, 83}

\bibitem[\protect\citeauthoryear{{Croton} et~al.,}{{Croton}
  et~al.}{2006}]{Croton2006MNRAS}
{Croton} D.~J.,  et~al., 2006, \mn@doi [\mnras]
  {10.1111/j.1365-2966.2005.09675.x}, \href
  {http://adsabs.harvard.edu/abs/2006MNRAS.365...11C} {365, 11}

\bibitem[\protect\citeauthoryear{{Cui}, {Yuan}  \& {Li}}{{Cui}
  et~al.}{2020}]{Cui2020ApJ}
{Cui} C.,  {Yuan} F.,   {Li} B.,  2020, \mn@doi [\apj]
  {10.3847/1538-4357/ab6e6e}, \href
  {https://ui.adsabs.harvard.edu/abs/2020ApJ...890...80C} {890, 80}

\bibitem[\protect\citeauthoryear{{Dasyra} \& {Combes}}{{Dasyra} \&
  {Combes}}{2011}]{Dasyra2011AA}
{Dasyra} K.~M.,  {Combes} F.,  2011, \mn@doi [\aap]
  {10.1051/0004-6361/201117730}, \href
  {https://ui.adsabs.harvard.edu/abs/2011A&A...533L..10D} {533, L10}

\bibitem[\protect\citeauthoryear{{Downes} \& {Solomon}}{{Downes} \&
  {Solomon}}{1998}]{Downes1998ApJ}
{Downes} D.,  {Solomon} P.~M.,  1998, \mn@doi [\apj] {10.1086/306339}, \href
  {https://ui.adsabs.harvard.edu/abs/1998ApJ...507..615D} {507, 615}

\bibitem[\protect\citeauthoryear{{Dubois} et~al.,}{{Dubois}
  et~al.}{2014}]{Dubois2014MNRAS}
{Dubois} Y.,  et~al., 2014, \mn@doi [\mnras] {10.1093/mnras/stu1227}, \href
  {https://ui.adsabs.harvard.edu/abs/2014MNRAS.444.1453D} {444, 1453}

\bibitem[\protect\citeauthoryear{{Faucher-Gigu{\`e}re} \&
  {Quataert}}{{Faucher-Gigu{\`e}re} \& {Quataert}}{2012}]{Faucher2012MNRASb}
{Faucher-Gigu{\`e}re} C.-A.,  {Quataert} E.,  2012, \mn@doi [\mnras]
  {10.1111/j.1365-2966.2012.21512.x}, \href
  {http://adsabs.harvard.edu/abs/2012MNRAS.425..605F} {425, 605}

\bibitem[\protect\citeauthoryear{{Feruglio}, {Maiolino}, {Piconcelli}, {Menci},
  {Aussel}, {Lamastra}  \& {Fiore}}{{Feruglio} et~al.}{2010}]{Feruglio2010AA}
{Feruglio} C.,  {Maiolino} R.,  {Piconcelli} E.,  {Menci} N.,  {Aussel} H.,
  {Lamastra} A.,   {Fiore} F.,  2010, \mn@doi [\aap]
  {10.1051/0004-6361/201015164}, \href
  {http://adsabs.harvard.edu/abs/2010A%26A...518L.155F} {518, L155+}

\bibitem[\protect\citeauthoryear{{Feruglio}, {Fiore}, {Piconcelli}, {Cicone},
  {Maiolino}, {Davies}  \& {Sturm}}{{Feruglio} et~al.}{2013}]{Feruglio2013AA}
{Feruglio} C.,  {Fiore} F.,  {Piconcelli} E.,  {Cicone} C.,  {Maiolino} R.,
  {Davies} R.,   {Sturm} E.,  2013, \mn@doi [\aap]
  {10.1051/0004-6361/201321275}, \href
  {https://ui.adsabs.harvard.edu/abs/2013A&A...558A..87F} {558, A87}

\bibitem[\protect\citeauthoryear{{Feruglio} et~al.,}{{Feruglio}
  et~al.}{2015}]{Feruglio2015AA}
{Feruglio} C.,  et~al., 2015, \mn@doi [\aap] {10.1051/0004-6361/201526020},
  \href {https://ui.adsabs.harvard.edu/abs/2015AA...583A..99F} {583, A99}

\bibitem[\protect\citeauthoryear{{Fiore} et~al.,}{{Fiore}
  et~al.}{2017}]{Fiore2017AA}
{Fiore} F.,  et~al., 2017, \mn@doi [\aap] {10.1051/0004-6361/201629478}, \href
  {http://adsabs.harvard.edu/abs/2017A%26A...601A.143F} {601, A143}

\bibitem[\protect\citeauthoryear{{Fluetsch} et~al.,}{{Fluetsch}
  et~al.}{2019}]{Fluetsch2019MNRAS}
{Fluetsch} A.,  et~al., 2019, \mn@doi [\mnras] {10.1093/mnras/sty3449}, \href
  {https://ui.adsabs.harvard.edu/abs/2019MNRAS.483.4586F} {483, 4586}

\bibitem[\protect\citeauthoryear{{Fukumura}, {Kazanas}, {Contopoulos}  \&
  {Behar}}{{Fukumura} et~al.}{2010}]{Fukumura2010ApJ}
{Fukumura} K.,  {Kazanas} D.,  {Contopoulos} I.,   {Behar} E.,  2010, \mn@doi
  [\apj] {10.1088/0004-637X/715/1/636}, \href
  {https://ui.adsabs.harvard.edu/abs/2010ApJ...715..636F} {715, 636}

\bibitem[\protect\citeauthoryear{{Garc{\'\i}a-Burillo}
  et~al.,}{{Garc{\'\i}a-Burillo} et~al.}{2014}]{GarciaB2014AA}
{Garc{\'\i}a-Burillo} S.,  et~al., 2014, \mn@doi [\aap]
  {10.1051/0004-6361/201423843}, \href
  {https://ui.adsabs.harvard.edu/abs/2014A&A...567A.125G} {567, A125}

\bibitem[\protect\citeauthoryear{{Garc{\'\i}a-Burillo}
  et~al.,}{{Garc{\'\i}a-Burillo} et~al.}{2015}]{GarciaB2015AA}
{Garc{\'\i}a-Burillo} S.,  et~al., 2015, \mn@doi [\aap]
  {10.1051/0004-6361/201526133}, \href
  {https://ui.adsabs.harvard.edu/abs/2015A&A...580A..35G} {580, A35}

\bibitem[\protect\citeauthoryear{{Giustini} \& {Proga}}{{Giustini} \&
  {Proga}}{2019}]{Giustini2019AA}
{Giustini} M.,  {Proga} D.,  2019, \mn@doi [\aap]
  {10.1051/0004-6361/201833810}, \href
  {https://ui.adsabs.harvard.edu/abs/2019AA...630A..94G} {630, A94}

\bibitem[\protect\citeauthoryear{{Gofford}, {Reeves}, {Tombesi}, {Braito},
  {Turner}, {Miller}  \& {Cappi}}{{Gofford} et~al.}{2013}]{Gofford2013MNRAS}
{Gofford} J.,  {Reeves} J.~N.,  {Tombesi} F.,  {Braito} V.,  {Turner} T.~J.,
  {Miller} L.,   {Cappi} M.,  2013, \mn@doi [\mnras] {10.1093/mnras/sts481},
  \href {https://ui.adsabs.harvard.edu/abs/2013MNRAS.430...60G} {430, 60}

\bibitem[\protect\citeauthoryear{{Gonz{\'a}lez-Alfonso}
  et~al.,}{{Gonz{\'a}lez-Alfonso} et~al.}{2017}]{Gonzalez2017ApJ}
{Gonz{\'a}lez-Alfonso} E.,  et~al., 2017, \mn@doi [\apj]
  {10.3847/1538-4357/836/1/11}, \href
  {https://ui.adsabs.harvard.edu/abs/2017ApJ...836...11G} {836, 11}

\bibitem[\protect\citeauthoryear{{Gowardhan} et~al.,}{{Gowardhan}
  et~al.}{2018}]{Gowardhan2018ApJ}
{Gowardhan} A.,  et~al., 2018, \mn@doi [\apj] {10.3847/1538-4357/aabccc}, \href
  {https://ui.adsabs.harvard.edu/abs/2018ApJ...859...35G} {859, 35}

\bibitem[\protect\citeauthoryear{{Grand} et~al.,}{{Grand}
  et~al.}{2017}]{Grand2017MNRAS}
{Grand} R. J.~J.,  et~al., 2017, \mn@doi [\mnras] {10.1093/mnras/stx071}, \href
  {https://ui.adsabs.harvard.edu/abs/2017MNRAS.467..179G} {467, 179}

\bibitem[\protect\citeauthoryear{{Hamann}, {Chartas}, {Reeves}  \&
  {Nardini}}{{Hamann} et~al.}{2018}]{Hamann2018MNRAS}
{Hamann} F.,  {Chartas} G.,  {Reeves} J.,   {Nardini} E.,  2018, \mn@doi
  [\mnras] {10.1093/mnras/sty043}, \href
  {https://ui.adsabs.harvard.edu/abs/2018MNRAS.476..943H} {476, 943}

\bibitem[\protect\citeauthoryear{{Harrison}, {Costa}, {Tadhunter},
  {Fl{\"u}tsch}, {Kakkad}, {Perna}  \& {Vietri}}{{Harrison}
  et~al.}{2018}]{Harrison2018NatAs}
{Harrison} C.~M.,  {Costa} T.,  {Tadhunter} C.~N.,  {Fl{\"u}tsch} A.,  {Kakkad}
  D.,  {Perna} M.,   {Vietri} G.,  2018, \mn@doi [Nature Astronomy]
  {10.1038/s41550-018-0403-6}, \href
  {http://adsabs.harvard.edu/abs/2018NatAs...2..198H} {2, 198}

\bibitem[\protect\citeauthoryear{{Heckman} \& {Best}}{{Heckman} \&
  {Best}}{2014}]{Heckman2014ARAA}
{Heckman} T.~M.,  {Best} P.~N.,  2014, \mn@doi [\araa]
  {10.1146/annurev-astro-081913-035722}, \href
  {http://adsabs.harvard.edu/abs/2014ARA%26A..52..589H} {52, 589}

\bibitem[\protect\citeauthoryear{{Higginbottom}, {Proga}, {Knigge}, {Long},
  {Matthews}  \& {Sim}}{{Higginbottom} et~al.}{2014}]{Higginbottom2014ApJ}
{Higginbottom} N.,  {Proga} D.,  {Knigge} C.,  {Long} K.~S.,  {Matthews} J.~H.,
    {Sim} S.~A.,  2014, \mn@doi [\apj] {10.1088/0004-637X/789/1/19}, \href
  {https://ui.adsabs.harvard.edu/abs/2014ApJ...789...19H} {789, 19}

\bibitem[\protect\citeauthoryear{{Hopkins} \& {Elvis}}{{Hopkins} \&
  {Elvis}}{2010}]{Hopkins2010MNRAS}
{Hopkins} P.~F.,  {Elvis} M.,  2010, \mn@doi [\mnras]
  {10.1111/j.1365-2966.2009.15643.x}, \href
  {http://adsabs.harvard.edu/abs/2010MNRAS.401....7H} {401, 7}

\bibitem[\protect\citeauthoryear{{Hopkins}, {Hernquist}, {Martini}, {Cox},
  {Robertson}, {Di Matteo}  \& {Springel}}{{Hopkins}
  et~al.}{2005}]{Hopkins2005ApJ}
{Hopkins} P.~F.,  {Hernquist} L.,  {Martini} P.,  {Cox} T.~J.,  {Robertson} B.,
   {Di Matteo} T.,   {Springel} V.,  2005, \mn@doi [\apjl] {10.1086/431146},
  \href {http://adsabs.harvard.edu/abs/2005ApJ...625L..71H} {625, L71}

\bibitem[\protect\citeauthoryear{{Ishibashi} \& {Fabian}}{{Ishibashi} \&
  {Fabian}}{2015}]{Ishibashi2015MNRAS}
{Ishibashi} W.,  {Fabian} A.~C.,  2015, \mn@doi [\mnras]
  {10.1093/mnras/stv944}, \href
  {https://ui.adsabs.harvard.edu/abs/2015MNRAS.451...93I} {451, 93}

\bibitem[\protect\citeauthoryear{{King}}{{King}}{2003}]{King2003ApJ}
{King} A.,  2003, \mn@doi [\apjl] {10.1086/379143}, \href
  {http://adsabs.harvard.edu/abs/2003ApJ...596L..27K} {596, L27}

\bibitem[\protect\citeauthoryear{{King}}{{King}}{2005}]{King2005ApJ}
{King} A.,  2005, \mn@doi [\apjl] {10.1086/499430}, \href
  {http://adsabs.harvard.edu/abs/2005ApJ...635L.121K} {635, L121}

\bibitem[\protect\citeauthoryear{{King}}{{King}}{2010a}]{King2010MNRASa}
{King} A.~R.,  2010a, \mn@doi [\mnras] {10.1111/j.1365-2966.2009.16013.x},
  \href {http://adsabs.harvard.edu/abs/2010MNRAS.402.1516K} {402, 1516}

\bibitem[\protect\citeauthoryear{{King}}{{King}}{2010b}]{King2010MNRASb}
{King} A.~R.,  2010b, \mn@doi [\mnras] {10.1111/j.1745-3933.2010.00938.x},
  \href {http://adsabs.harvard.edu/abs/2010MNRAS.408L..95K} {408, L95}

\bibitem[\protect\citeauthoryear{{King} \& {Nixon}}{{King} \&
  {Nixon}}{2015}]{King2015MNRAS}
{King} A.,  {Nixon} C.,  2015, \mn@doi [\mnras] {10.1093/mnrasl/slv098}, \href
  {http://adsabs.harvard.edu/abs/2015MNRAS.453L..46K} {453, L46}

\bibitem[\protect\citeauthoryear{{King} \& {Pounds}}{{King} \&
  {Pounds}}{2003}]{King2003MNRASb}
{King} A.~R.,  {Pounds} K.~A.,  2003, \mn@doi [\mnras]
  {10.1046/j.1365-8711.2003.06980.x}, \href
  {http://adsabs.harvard.edu/abs/2003MNRAS.345..657K} {345, 657}

\bibitem[\protect\citeauthoryear{{King} \& {Pounds}}{{King} \&
  {Pounds}}{2015}]{King2015ARAA}
{King} A.,  {Pounds} K.,  2015, \mn@doi [\araa]
  {10.1146/annurev-astro-082214-122316}, \href
  {http://adsabs.harvard.edu/abs/2015ARA%26A..53..115K} {53, 115}

\bibitem[\protect\citeauthoryear{{King} \& {Pringle}}{{King} \&
  {Pringle}}{2007}]{King2007MNRAS}
{King} A.~R.,  {Pringle} J.~E.,  2007, \mn@doi [\mnras]
  {10.1111/j.1745-3933.2007.00296.x}, \href
  {http://adsabs.harvard.edu/abs/2007MNRAS.377L..25K} {377, L25}

\bibitem[\protect\citeauthoryear{{King}, {Zubovas}  \& {Power}}{{King}
  et~al.}{2011}]{King2011MNRAS}
{King} A.~R.,  {Zubovas} K.,   {Power} C.,  2011, \mn@doi [\mnras]
  {10.1111/j.1745-3933.2011.01067.x}, \href
  {http://adsabs.harvard.edu/abs/2011MNRAS.415L...6K} {415, L6}

\bibitem[\protect\citeauthoryear{{Larson}}{{Larson}}{1981}]{Larson1981MNRAS}
{Larson} R.~B.,  1981, \mnras, \href
  {http://adsabs.harvard.edu/abs/1981MNRAS.194..809L} {194, 809}

\bibitem[\protect\citeauthoryear{{Leon} et~al.,}{{Leon}
  et~al.}{2007}]{Leon2007AA}
{Leon} S.,  et~al., 2007, \mn@doi [\aap] {10.1051/0004-6361:20066075}, \href
  {https://ui.adsabs.harvard.edu/abs/2007A&A...473..747L} {473, 747}

\bibitem[\protect\citeauthoryear{{Lutz} et~al.,}{{Lutz}
  et~al.}{2020}]{Lutz2020AA}
{Lutz} D.,  et~al., 2020, \mn@doi [\aap] {10.1051/0004-6361/201936803}, \href
  {https://ui.adsabs.harvard.edu/abs/2020AA...633A.134L} {633, A134}

\bibitem[\protect\citeauthoryear{{Maiolino} et~al.,}{{Maiolino}
  et~al.}{2012}]{Maiolino2012MNRAS}
{Maiolino} R.,  et~al., 2012, \mn@doi [\mnras]
  {10.1111/j.1745-3933.2012.01303.x}, \href
  {https://ui.adsabs.harvard.edu/abs/2012MNRAS.425L..66M} {425, L66}

\bibitem[\protect\citeauthoryear{{Matthews}, {Knigge}, {Long}, {Sim},
  {Higginbottom}  \& {Mangham}}{{Matthews} et~al.}{2016}]{Matthews2016MNRAS}
{Matthews} J.~H.,  {Knigge} C.,  {Long} K.~S.,  {Sim} S.~A.,  {Higginbottom}
  N.,   {Mangham} S.~W.,  2016, \mn@doi [\mnras] {10.1093/mnras/stw323}, \href
  {https://ui.adsabs.harvard.edu/abs/2016MNRAS.458..293M} {458, 293}

\bibitem[\protect\citeauthoryear{{Matzeu}, {Reeves}, {Nardini}, {Braito},
  {Costa}, {Tombesi}  \& {Gofford}}{{Matzeu} et~al.}{2016}]{Matzeu2016MNRAS}
{Matzeu} G.~A.,  {Reeves} J.~N.,  {Nardini} E.,  {Braito} V.,  {Costa} M.~T.,
  {Tombesi} F.,   {Gofford} J.,  2016, \mn@doi [\mnras] {10.1093/mnras/stw354},
  \href {https://ui.adsabs.harvard.edu/abs/2016MNRAS.458.1311M} {458, 1311}

\bibitem[\protect\citeauthoryear{{Matzeu}, {Reeves}, {Braito}, {Nardini},
  {McLaughlin}, {Lobban}, {Tombesi}  \& {Costa}}{{Matzeu}
  et~al.}{2017}]{Matzeu2017MNRAS}
{Matzeu} G.~A.,  {Reeves} J.~N.,  {Braito} V.,  {Nardini} E.,  {McLaughlin}
  D.~E.,  {Lobban} A.~P.,  {Tombesi} F.,   {Costa} M.~T.,  2017, \mn@doi
  [\mnras] {10.1093/mnrasl/slx129}, \href
  {https://ui.adsabs.harvard.edu/abs/2017MNRAS.472L..15M} {472, L15}

\bibitem[\protect\citeauthoryear{{McConnell} \& {Ma}}{{McConnell} \&
  {Ma}}{2013}]{McConnell2013ApJ}
{McConnell} N.~J.,  {Ma} C.-P.,  2013, \mn@doi [\apj]
  {10.1088/0004-637X/764/2/184}, \href
  {http://adsabs.harvard.edu/abs/2013ApJ...764..184M} {764, 184}

\bibitem[\protect\citeauthoryear{{Morganti}, {Frieswijk}, {Oonk}, {Oosterloo}
  \& {Tadhunter}}{{Morganti} et~al.}{2013}]{Morganti2013AA}
{Morganti} R.,  {Frieswijk} W.,  {Oonk} R.~J.~B.,  {Oosterloo} T.,
  {Tadhunter} C.,  2013, \mn@doi [\aap] {10.1051/0004-6361/201220734}, \href
  {https://ui.adsabs.harvard.edu/abs/2013A&A...552L...4M} {552, L4}

\bibitem[\protect\citeauthoryear{{Murray}, {Quataert}  \& {Thompson}}{{Murray}
  et~al.}{2005}]{Murray2005ApJ}
{Murray} N.,  {Quataert} E.,   {Thompson} T.~A.,  2005, \mn@doi [\apj]
  {10.1086/426067}, \href {http://adsabs.harvard.edu/abs/2005ApJ...618..569M}
  {618, 569}

\bibitem[\protect\citeauthoryear{{Nardini} \& {Zubovas}}{{Nardini} \&
  {Zubovas}}{2018}]{Nardini2018MNRAS}
{Nardini} E.,  {Zubovas} K.,  2018, \mn@doi [\mnras] {10.1093/mnras/sty1144},
  \href {http://adsabs.harvard.edu/abs/2018MNRAS.tmp.1081N} {}

\bibitem[\protect\citeauthoryear{{Nardini} et~al.,}{{Nardini}
  et~al.}{2015}]{Nardini2015Sci}
{Nardini} E.,  et~al., 2015, \mn@doi [Science] {10.1126/science.1259202}, \href
  {http://adsabs.harvard.edu/abs/2015Sci...347..860N} {347, 860}

\bibitem[\protect\citeauthoryear{{Navarro}, {Frenk}  \& {White}}{{Navarro}
  et~al.}{1997}]{Navarro1997ApJ}
{Navarro} J.~F.,  {Frenk} C.~S.,   {White} S.~D.~M.,  1997, \mn@doi [\apj]
  {10.1086/304888}, \href {http://adsabs.harvard.edu/abs/1997ApJ...490..493N}
  {490, 493}

\bibitem[\protect\citeauthoryear{{Nelson} et~al.,}{{Nelson}
  et~al.}{2019}]{Nelson2019MNRAS}
{Nelson} D.,  et~al., 2019, \mn@doi [\mnras] {10.1093/mnras/stz2306}, \href
  {https://ui.adsabs.harvard.edu/abs/2019MNRAS.490.3234N} {490, 3234}

\bibitem[\protect\citeauthoryear{{Nims}, {Quataert}  \&
  {Faucher-Gigu{\`e}re}}{{Nims} et~al.}{2015}]{Nims2015MNRAS}
{Nims} J.,  {Quataert} E.,   {Faucher-Gigu{\`e}re} C.-A.,  2015, \mn@doi
  [\mnras] {10.1093/mnras/stu2648}, \href
  {https://ui.adsabs.harvard.edu/abs/2015MNRAS.447.3612N} {447, 3612}

\bibitem[\protect\citeauthoryear{{Nomura}, {Ohsuga}, {Takahashi}, {Wada}  \&
  {Yoshida}}{{Nomura} et~al.}{2016}]{Nomura2016PASJ}
{Nomura} M.,  {Ohsuga} K.,  {Takahashi} H.~R.,  {Wada} K.,   {Yoshida} T.,
  2016, \mn@doi [\pasj] {10.1093/pasj/psv124}, \href
  {https://ui.adsabs.harvard.edu/abs/2016PASJ...68...16N} {68, 16}

\bibitem[\protect\citeauthoryear{{Papadopoulos}, {van der Werf}, {Xilouris},
  {Isaak}  \& {Gao}}{{Papadopoulos} et~al.}{2012}]{Papadopoulos2012ApJ}
{Papadopoulos} P.~P.,  {van der Werf} P.,  {Xilouris} E.,  {Isaak} K.~G.,
  {Gao} Y.,  2012, \mn@doi [\apj] {10.1088/0004-637X/751/1/10}, \href
  {https://ui.adsabs.harvard.edu/abs/2012ApJ...751...10P} {751, 10}

\bibitem[\protect\citeauthoryear{{Pereira-Santaella}
  et~al.,}{{Pereira-Santaella} et~al.}{2016}]{PereiraS2016AA}
{Pereira-Santaella} M.,  et~al., 2016, \mn@doi [\aap]
  {10.1051/0004-6361/201628875}, \href
  {https://ui.adsabs.harvard.edu/abs/2016A&A...594A..81P} {594, A81}

\bibitem[\protect\citeauthoryear{{Pereira-Santaella}
  et~al.,}{{Pereira-Santaella} et~al.}{2018}]{PereiraS2018AA}
{Pereira-Santaella} M.,  et~al., 2018, \mn@doi [\aap]
  {10.1051/0004-6361/201833089}, \href
  {https://ui.adsabs.harvard.edu/abs/2018A&A...616A.171P} {616, A171}

\bibitem[\protect\citeauthoryear{{Pounds} \& {King}}{{Pounds} \&
  {King}}{2013}]{Pounds2013MNRAS}
{Pounds} K.~A.,  {King} A.~R.,  2013, \mn@doi [\mnras] {10.1093/mnras/stt807},
  \href {http://adsabs.harvard.edu/abs/2013MNRAS.433.1369P} {433, 1369}

\bibitem[\protect\citeauthoryear{{Pounds}, {Reeves}, {King}, {Page}, {O'Brien}
  \& {Turner}}{{Pounds} et~al.}{2003}]{Pounds2003MNRASa}
{Pounds} K.~A.,  {Reeves} J.~N.,  {King} A.~R.,  {Page} K.~L.,  {O'Brien}
  P.~T.,   {Turner} M.~J.~L.,  2003, \mn@doi [\mnras]
  {10.1046/j.1365-8711.2003.07006.x}, \href
  {http://adsabs.harvard.edu/abs/2003MNRAS.345..705P} {345, 705}

\bibitem[\protect\citeauthoryear{{Pounds}, {Lobban}, {Reeves}  \&
  {Vaughan}}{{Pounds} et~al.}{2016}]{Pounds2016MNRAS}
{Pounds} K.,  {Lobban} A.,  {Reeves} J.,   {Vaughan} S.,  2016, \mn@doi
  [\mnras] {10.1093/mnras/stw165}, \href
  {https://ui.adsabs.harvard.edu/abs/2016MNRAS.457.2951P} {457, 2951}

\bibitem[\protect\citeauthoryear{{Power}, {Nayakshin}  \& {King}}{{Power}
  et~al.}{2011}]{Power2011MNRASb}
{Power} C.,  {Nayakshin} S.,   {King} A.,  2011, \mn@doi [\mnras]
  {10.1111/j.1365-2966.2010.17901.x}, \href
  {http://adsabs.harvard.edu/abs/2011MNRAS.412..269P} {412, 269}

\bibitem[\protect\citeauthoryear{{Proga}}{{Proga}}{2003}]{Proga2003ApJ}
{Proga} D.,  2003, \mn@doi [\apj] {10.1086/345897}, \href
  {https://ui.adsabs.harvard.edu/abs/2003ApJ...585..406P} {585, 406}

\bibitem[\protect\citeauthoryear{{Proga}, {Stone}  \& {Kallman}}{{Proga}
  et~al.}{2000}]{Proga2000ApJ}
{Proga} D.,  {Stone} J.~M.,   {Kallman} T.~R.,  2000, \mn@doi [\apj]
  {10.1086/317154}, \href {http://adsabs.harvard.edu/abs/2000ApJ...543..686P}
  {543, 686}

\bibitem[\protect\citeauthoryear{{Puchwein} \& {Springel}}{{Puchwein} \&
  {Springel}}{2013}]{Puchwein2013MNRAS}
{Puchwein} E.,  {Springel} V.,  2013, \mn@doi [\mnras] {10.1093/mnras/sts243},
  \href {https://ui.adsabs.harvard.edu/abs/2013MNRAS.428.2966P} {428, 2966}

\bibitem[\protect\citeauthoryear{{Querejeta} et~al.,}{{Querejeta}
  et~al.}{2016}]{Querejeta2016AA}
{Querejeta} M.,  et~al., 2016, \mn@doi [\aap] {10.1051/0004-6361/201628674},
  \href {https://ui.adsabs.harvard.edu/abs/2016A&A...593A.118Q} {593, A118}

\bibitem[\protect\citeauthoryear{{Reeves} \& {Braito}}{{Reeves} \&
  {Braito}}{2019}]{Reeves2019ApJ}
{Reeves} J.~N.,  {Braito} V.,  2019, \mn@doi [\apj] {10.3847/1538-4357/ab41f9},
  \href {https://ui.adsabs.harvard.edu/abs/2019ApJ...884...80R} {884, 80}

\bibitem[\protect\citeauthoryear{{Reeves}, {O'Brien}  \& {Ward}}{{Reeves}
  et~al.}{2003}]{Reeves2003ApJ}
{Reeves} J.~N.,  {O'Brien} P.~T.,   {Ward} M.~J.,  2003, \mn@doi [\apjl]
  {10.1086/378218}, \href
  {https://ui.adsabs.harvard.edu/abs/2003ApJ...593L..65R} {593, L65}

\bibitem[\protect\citeauthoryear{{Reeves}, {Lobban}  \& {Pounds}}{{Reeves}
  et~al.}{2018a}]{Reeves2018ApJa}
{Reeves} J.~N.,  {Lobban} A.,   {Pounds} K.~A.,  2018a, \mn@doi [\apj]
  {10.3847/1538-4357/aaa776}, \href
  {https://ui.adsabs.harvard.edu/abs/2018ApJ...854...28R} {854, 28}

\bibitem[\protect\citeauthoryear{{Reeves}, {Braito}, {Nardini}, {Lobban},
  {Matzeu}  \& {Costa}}{{Reeves} et~al.}{2018b}]{Reeves2018ApJb}
{Reeves} J.~N.,  {Braito} V.,  {Nardini} E.,  {Lobban} A.~P.,  {Matzeu} G.~A.,
   {Costa} M.~T.,  2018b, \mn@doi [\apjl] {10.3847/2041-8213/aaaae1}, \href
  {https://ui.adsabs.harvard.edu/abs/2018ApJ...854L...8R} {854, L8}

\bibitem[\protect\citeauthoryear{{Reeves}, {Braito}, {Chartas}, {Hamann},
  {Laha}  \& {Nardini}}{{Reeves} et~al.}{2020}]{Reeves2020arXiv}
{Reeves} J.,  {Braito} V.,  {Chartas} G.,  {Hamann} F.,  {Laha} S.,   {Nardini}
  E.,  2020, arXiv e-prints, \href
  {https://ui.adsabs.harvard.edu/abs/2020arXiv200412439R} {p. arXiv:2004.12439}

\bibitem[\protect\citeauthoryear{{Richards} et~al.,}{{Richards}
  et~al.}{2006}]{Richards2006ApJS}
{Richards} G.~T.,  et~al., 2006, \mn@doi [\apjs] {10.1086/506525}, \href
  {https://ui.adsabs.harvard.edu/abs/2006ApJS..166..470R} {166, 470}

\bibitem[\protect\citeauthoryear{{Richings} \&
  {Faucher-Gigu{\`e}re}}{{Richings} \&
  {Faucher-Gigu{\`e}re}}{2018a}]{Richings2018MNRAS}
{Richings} A.~J.,  {Faucher-Gigu{\`e}re} C.-A.,  2018a, \mn@doi [\mnras]
  {10.1093/mnras/stx3014}, \href
  {http://adsabs.harvard.edu/abs/2018MNRAS.474.3673R} {474, 3673}

\bibitem[\protect\citeauthoryear{{Richings} \&
  {Faucher-Gigu{\`e}re}}{{Richings} \&
  {Faucher-Gigu{\`e}re}}{2018b}]{Richings2018MNRASb}
{Richings} A.~J.,  {Faucher-Gigu{\`e}re} C.-A.,  2018b, \mn@doi [\mnras]
  {10.1093/mnras/sty1285}, \href
  {http://adsabs.harvard.edu/abs/2018MNRAS.478.3100R} {478, 3100}

\bibitem[\protect\citeauthoryear{{Roberts-Borsani}}{{Roberts-Borsani}}{2020}]{RobertsBorsani2020MNRAS}
{Roberts-Borsani} G.~W.,  2020, \mn@doi [\mnras] {10.1093/mnras/staa1006},
  \href {https://ui.adsabs.harvard.edu/abs/2020MNRAS.tmp.1154R} {}

\bibitem[\protect\citeauthoryear{{Rupke} \& {Veilleux}}{{Rupke} \&
  {Veilleux}}{2011}]{Rupke2011ApJ}
{Rupke} D.~S.~N.,  {Veilleux} S.,  2011, \mn@doi [\apjl]
  {10.1088/2041-8205/729/2/L27}, \href
  {http://adsabs.harvard.edu/abs/2011ApJ...729L..27R} {729, L27+}

\bibitem[\protect\citeauthoryear{{Rupke}, {Veilleux}  \& {Sanders}}{{Rupke}
  et~al.}{2005}]{Rupke2005ApJS}
{Rupke} D.~S.,  {Veilleux} S.,   {Sanders} D.~B.,  2005, \mn@doi [\apjs]
  {10.1086/432889}, \href
  {https://ui.adsabs.harvard.edu/abs/2005ApJS..160..115R} {160, 115}

\bibitem[\protect\citeauthoryear{{Rupke}, {G{\"u}ltekin}  \&
  {Veilleux}}{{Rupke} et~al.}{2017}]{Rupke2017ApJ}
{Rupke} D. S.~N.,  {G{\"u}ltekin} K.,   {Veilleux} S.,  2017, \mn@doi [\apj]
  {10.3847/1538-4357/aa94d1}, \href
  {https://ui.adsabs.harvard.edu/abs/2017ApJ...850...40R} {850, 40}

\bibitem[\protect\citeauthoryear{{Sakamoto}, {Aalto}, {Combes}, {Evans}  \&
  {Peck}}{{Sakamoto} et~al.}{2014}]{Sakamoto2014ApJ}
{Sakamoto} K.,  {Aalto} S.,  {Combes} F.,  {Evans} A.,   {Peck} A.,  2014,
  \mn@doi [\apj] {10.1088/0004-637X/797/2/90}, \href
  {https://ui.adsabs.harvard.edu/abs/2014ApJ...797...90S} {797, 90}

\bibitem[\protect\citeauthoryear{{Salak}, {Nakai}, {Hatakeyama}  \&
  {Miyamoto}}{{Salak} et~al.}{2016}]{Salak2016ApJ}
{Salak} D.,  {Nakai} N.,  {Hatakeyama} T.,   {Miyamoto} Y.,  2016, \mn@doi
  [\apj] {10.3847/0004-637X/823/1/68}, \href
  {https://ui.adsabs.harvard.edu/abs/2016ApJ...823...68S} {823, 68}

\bibitem[\protect\citeauthoryear{{Schawinski}, {Koss}, {Berney}  \&
  {Sartori}}{{Schawinski} et~al.}{2015}]{Schawinski2015MNRAS}
{Schawinski} K.,  {Koss} M.,  {Berney} S.,   {Sartori} L.~F.,  2015, \mn@doi
  [\mnras] {10.1093/mnras/stv1136}, \href
  {http://adsabs.harvard.edu/abs/2015MNRAS.451.2517S} {451, 2517}

\bibitem[\protect\citeauthoryear{{Schaye} et~al.,}{{Schaye}
  et~al.}{2015}]{Schaye2015MNRAS}
{Schaye} J.,  et~al., 2015, \mn@doi [\mnras] {10.1093/mnras/stu2058}, \href
  {http://adsabs.harvard.edu/abs/2015MNRAS.446..521S} {446, 521}

\bibitem[\protect\citeauthoryear{{Shankar}, {Weinberg}  \&
  {Miralda-Escud{\'e}}}{{Shankar} et~al.}{2013}]{Shankar2013MNRAS}
{Shankar} F.,  {Weinberg} D.~H.,   {Miralda-Escud{\'e}} J.,  2013, \mn@doi
  [\mnras] {10.1093/mnras/sts026}, \href
  {http://adsabs.harvard.edu/abs/2013MNRAS.428..421S} {428, 421}

\bibitem[\protect\citeauthoryear{{Sijacki}, {Springel}, {Di Matteo}  \&
  {Hernquist}}{{Sijacki} et~al.}{2007}]{Sijacki2007MNRAS}
{Sijacki} D.,  {Springel} V.,  {Di Matteo} T.,   {Hernquist} L.,  2007, \mn@doi
  [\mnras] {10.1111/j.1365-2966.2007.12153.x}, \href
  {http://adsabs.harvard.edu/abs/2007MNRAS.380..877S} {380, 877}

\bibitem[\protect\citeauthoryear{{Sim}, {Long}, {Miller}  \& {Turner}}{{Sim}
  et~al.}{2008}]{Sim2008MNRAS}
{Sim} S.~A.,  {Long} K.~S.,  {Miller} L.,   {Turner} T.~J.,  2008, \mn@doi
  [\mnras] {10.1111/j.1365-2966.2008.13466.x}, \href
  {https://ui.adsabs.harvard.edu/abs/2008MNRAS.388..611S} {388, 611}

\bibitem[\protect\citeauthoryear{{Simpson}, {Ward}, {O'Brien}  \&
  {Reeves}}{{Simpson} et~al.}{1999}]{Simpson1999MNRAS}
{Simpson} C.,  {Ward} M.,  {O'Brien} P.,   {Reeves} J.,  1999, \mn@doi [\mnras]
  {10.1046/j.1365-8711.1999.02344.x}, \href
  {https://ui.adsabs.harvard.edu/abs/1999MNRAS.303L..23S} {303, L23}

\bibitem[\protect\citeauthoryear{{Sirressi} et~al.,}{{Sirressi}
  et~al.}{2019}]{Sirressi2019MNRAS}
{Sirressi} M.,  et~al., 2019, \mn@doi [\mnras] {10.1093/mnras/stz2249}, \href
  {https://ui.adsabs.harvard.edu/abs/2019MNRAS.tmp.2170S} {p.~2170}

\bibitem[\protect\citeauthoryear{{S{\k{a}}dowski}, {Narayan}, {Penna}  \&
  {Zhu}}{{S{\k{a}}dowski} et~al.}{2013}]{Sadowski2013MNRAS}
{S{\k{a}}dowski} A.,  {Narayan} R.,  {Penna} R.,   {Zhu} Y.,  2013, \mn@doi
  [\mnras] {10.1093/mnras/stt1881}, \href
  {https://ui.adsabs.harvard.edu/abs/2013MNRAS.436.3856S} {436, 3856}

\bibitem[\protect\citeauthoryear{{Smith}, {Tombesi}, {Veilleux}, {Lohfink}  \&
  {Luminari}}{{Smith} et~al.}{2019}]{Smith2019ApJ}
{Smith} R.~N.,  {Tombesi} F.,  {Veilleux} S.,  {Lohfink} A.~M.,   {Luminari}
  A.,  2019, \mn@doi [\apj] {10.3847/1538-4357/ab4ef8}, \href
  {https://ui.adsabs.harvard.edu/abs/2019ApJ...887...69S} {887, 69}

\bibitem[\protect\citeauthoryear{{Sturm} et~al.,}{{Sturm}
  et~al.}{2011}]{Sturm2011ApJ}
{Sturm} E.,  et~al., 2011, \mn@doi [\apjl] {10.1088/2041-8205/733/1/L16}, \href
  {http://adsabs.harvard.edu/abs/2011ApJ...733L..16S} {733, L16+}

\bibitem[\protect\citeauthoryear{{Sun}, {Greene}, {Zakamska}  \&
  {Nesvadba}}{{Sun} et~al.}{2014}]{Sun2014ApJ}
{Sun} A.-L.,  {Greene} J.~E.,  {Zakamska} N.~L.,   {Nesvadba} N. P.~H.,  2014,
  \mn@doi [\apj] {10.1088/0004-637X/790/2/160}, \href
  {https://ui.adsabs.harvard.edu/abs/2014ApJ...790..160S} {790, 160}

\bibitem[\protect\citeauthoryear{{Tombesi}, {Cappi}, {Reeves}, {Palumbo},
  {Yaqoob}, {Braito}  \& {Dadina}}{{Tombesi} et~al.}{2010a}]{Tombesi2010AA}
{Tombesi} F.,  {Cappi} M.,  {Reeves} J.~N.,  {Palumbo} G.~G.~C.,  {Yaqoob} T.,
  {Braito} V.,   {Dadina} M.,  2010a, \mn@doi [\aap]
  {10.1051/0004-6361/200913440}, \href
  {http://adsabs.harvard.edu/abs/2010A%26A...521A..57T} {521, A57+}

\bibitem[\protect\citeauthoryear{{Tombesi}, {Sambruna}, {Reeves}, {Braito},
  {Ballo}, {Gofford}, {Cappi}  \& {Mushotzky}}{{Tombesi}
  et~al.}{2010b}]{Tombesi2010ApJ}
{Tombesi} F.,  {Sambruna} R.~M.,  {Reeves} J.~N.,  {Braito} V.,  {Ballo} L.,
  {Gofford} J.,  {Cappi} M.,   {Mushotzky} R.~F.,  2010b, \mn@doi [\apj]
  {10.1088/0004-637X/719/1/700}, \href
  {http://adsabs.harvard.edu/abs/2010ApJ...719..700T} {719, 700}

\bibitem[\protect\citeauthoryear{{Tombesi}, {Cappi}, {Reeves}  \&
  {Braito}}{{Tombesi} et~al.}{2012}]{Tombesi2012MNRAS}
{Tombesi} F.,  {Cappi} M.,  {Reeves} J.~N.,   {Braito} V.,  2012, \mn@doi
  [\mnras] {10.1111/j.1745-3933.2012.01221.x}, \href
  {http://adsabs.harvard.edu/abs/2012MNRAS.422L...1T} {422, L1}

\bibitem[\protect\citeauthoryear{{Tombesi}, {Cappi}, {Reeves}, {Nemmen},
  {Braito}, {Gaspari}  \& {Reynolds}}{{Tombesi}
  et~al.}{2013}]{Tombesi2013MNRAS}
{Tombesi} F.,  {Cappi} M.,  {Reeves} J.~N.,  {Nemmen} R.~S.,  {Braito} V.,
  {Gaspari} M.,   {Reynolds} C.~S.,  2013, \mn@doi [\mnras]
  {10.1093/mnras/sts692}, \href
  {http://adsabs.harvard.edu/abs/2013MNRAS.430.1102T} {430, 1102}

\bibitem[\protect\citeauthoryear{{Tombesi}, {Tazaki}, {Mushotzky}, {Ueda},
  {Cappi}, {Gofford}, {Reeves}  \& {Guainazzi}}{{Tombesi}
  et~al.}{2014}]{Tombesi2014MNRAS}
{Tombesi} F.,  {Tazaki} F.,  {Mushotzky} R.~F.,  {Ueda} Y.,  {Cappi} M.,
  {Gofford} J.,  {Reeves} J.~N.,   {Guainazzi} M.,  2014, \mn@doi [\mnras]
  {10.1093/mnras/stu1297}, \href
  {http://adsabs.harvard.edu/abs/2014MNRAS.443.2154T} {443, 2154}

\bibitem[\protect\citeauthoryear{{Tombesi}, {Mel{\'e}ndez}, {Veilleux},
  {Reeves}, {Gonz{\'a}lez-Alfonso}  \& {Reynolds}}{{Tombesi}
  et~al.}{2015}]{Tombesi2015Natur}
{Tombesi} F.,  {Mel{\'e}ndez} M.,  {Veilleux} S.,  {Reeves} J.~N.,
  {Gonz{\'a}lez-Alfonso} E.,   {Reynolds} C.~S.,  2015, \mn@doi [\nat]
  {10.1038/nature14261}, \href
  {http://adsabs.harvard.edu/abs/2015Natur.519..436T} {519, 436}

\bibitem[\protect\citeauthoryear{{Torres}, {Quast}, {Coziol}, {Jablonski}, {de
  la Reza}, {L{\'e}pine}  \& {Greg{\'o}rio-Hetem}}{{Torres}
  et~al.}{1997}]{Torres1997ApJ}
{Torres} C. A.~O.,  {Quast} G.~R.,  {Coziol} R.,  {Jablonski} F.,  {de la Reza}
  R.,  {L{\'e}pine} J.~R.~D.,   {Greg{\'o}rio-Hetem} J.,  1997, \mn@doi [\apjl]
  {10.1086/310913}, \href
  {https://ui.adsabs.harvard.edu/abs/1997ApJ...488L..19T} {488, L19}

\bibitem[\protect\citeauthoryear{{Tremmel} et~al.,}{{Tremmel}
  et~al.}{2019}]{Tremmel2019MNRAS}
{Tremmel} M.,  et~al., 2019, \mn@doi [\mnras] {10.1093/mnras/sty3336}, \href
  {https://ui.adsabs.harvard.edu/abs/2019MNRAS.483.3336T} {483, 3336}

\bibitem[\protect\citeauthoryear{{Tsai} et~al.,}{{Tsai}
  et~al.}{2009}]{Tsai2009PASJ}
{Tsai} A.-L.,  et~al., 2009, \mn@doi [\pasj] {10.1093/pasj/61.2.237}, \href
  {https://ui.adsabs.harvard.edu/abs/2009PASJ...61..237T} {61, 237}

\bibitem[\protect\citeauthoryear{{Tsai}, {Matsushita}, {Kong}, {Matsumoto}  \&
  {Kohno}}{{Tsai} et~al.}{2012}]{Tsai2012ApJ}
{Tsai} A.-L.,  {Matsushita} S.,  {Kong} A. K.~H.,  {Matsumoto} H.,   {Kohno}
  K.,  2012, \mn@doi [\apj] {10.1088/0004-637X/752/1/38}, \href
  {https://ui.adsabs.harvard.edu/abs/2012ApJ...752...38T} {752, 38}

\bibitem[\protect\citeauthoryear{{Veilleux} et~al.,}{{Veilleux}
  et~al.}{2009}]{Veilleux2009ApJS}
{Veilleux} S.,  et~al., 2009, \mn@doi [\apjs] {10.1088/0067-0049/182/2/628},
  \href {http://adsabs.harvard.edu/abs/2009ApJS..182..628V} {182, 628}

\bibitem[\protect\citeauthoryear{{Veilleux}, {Bolatto}, {Tombesi},
  {Mel{\'e}ndez}, {Sturm}, {Gonz{\'a}lez-Alfonso}, {Fischer}  \&
  {Rupke}}{{Veilleux} et~al.}{2017}]{Veilleux2017ApJ}
{Veilleux} S.,  {Bolatto} A.,  {Tombesi} F.,  {Mel{\'e}ndez} M.,  {Sturm} E.,
  {Gonz{\'a}lez-Alfonso} E.,  {Fischer} J.,   {Rupke} D.~S.~N.,  2017, \mn@doi
  [\apj] {10.3847/1538-4357/aa767d}, \href
  {http://adsabs.harvard.edu/abs/2017ApJ...843...18V} {843, 18}

\bibitem[\protect\citeauthoryear{{Venturi} et~al.,}{{Venturi}
  et~al.}{2018}]{Venturi2018AA}
{Venturi} G.,  et~al., 2018, \mn@doi [\aap] {10.1051/0004-6361/201833668},
  \href {https://ui.adsabs.harvard.edu/abs/2018AA...619A..74V} {619, A74}

\bibitem[\protect\citeauthoryear{{Vogelsberger} et~al.,}{{Vogelsberger}
  et~al.}{2014}]{Vogelsberger2014MNRAS}
{Vogelsberger} M.,  et~al., 2014, \mn@doi [\mnras] {10.1093/mnras/stu1536},
  \href {http://adsabs.harvard.edu/abs/2014MNRAS.444.1518V} {444, 1518}

\bibitem[\protect\citeauthoryear{{Walter}, {Weiss}  \& {Scoville}}{{Walter}
  et~al.}{2002}]{Walter2002ApJ}
{Walter} F.,  {Weiss} A.,   {Scoville} N.,  2002, \mn@doi [\apjl]
  {10.1086/345287}, \href
  {https://ui.adsabs.harvard.edu/abs/2002ApJ...580L..21W} {580, L21}

\bibitem[\protect\citeauthoryear{{Wang}, {Chen}  \& {Zhang}}{{Wang}
  et~al.}{2006}]{Wang2006ApJ}
{Wang} J.-M.,  {Chen} Y.-M.,   {Zhang} F.,  2006, \mn@doi [\apjl]
  {10.1086/507271}, \href
  {https://ui.adsabs.harvard.edu/abs/2006ApJ...647L..17W} {647, L17}

\bibitem[\protect\citeauthoryear{{Wurster} \& {Thacker}}{{Wurster} \&
  {Thacker}}{2013}]{Wurster2013MNRAS}
{Wurster} J.,  {Thacker} R.~J.,  2013, \mn@doi [\mnras] {10.1093/mnras/stt182},
  \href {https://ui.adsabs.harvard.edu/abs/2013MNRAS.431..539W} {431, 539}

\bibitem[\protect\citeauthoryear{{Yu} \& {Tremaine}}{{Yu} \&
  {Tremaine}}{2002}]{Yu2002MNRAS}
{Yu} Q.,  {Tremaine} S.,  2002, \mn@doi [\mnras]
  {10.1046/j.1365-8711.2002.05532.x}, \href
  {http://adsabs.harvard.edu/abs/2002MNRAS.335..965Y} {335, 965}

\bibitem[\protect\citeauthoryear{{Zheng}, {Xia}, {Mao}, {Wu}  \&
  {Deng}}{{Zheng} et~al.}{2002}]{Zheng2002AJ}
{Zheng} X.~Z.,  {Xia} X.~Y.,  {Mao} S.,  {Wu} H.,   {Deng} Z.~G.,  2002,
  \mn@doi [\aj] {10.1086/340964}, \href
  {https://ui.adsabs.harvard.edu/abs/2002AJ....124...18Z} {124, 18}

\bibitem[\protect\citeauthoryear{{Zschaechner} et~al.,}{{Zschaechner}
  et~al.}{2016}]{Zschaechner2016ApJ}
{Zschaechner} L.~K.,  et~al., 2016, \mn@doi [\apj]
  {10.3847/0004-637X/832/2/142}, \href
  {https://ui.adsabs.harvard.edu/abs/2016ApJ...832..142Z} {832, 142}

\bibitem[\protect\citeauthoryear{{Zubovas}}{{Zubovas}}{2018}]{Zubovas2018MNRAS}
{Zubovas} K.,  2018, \mn@doi [\mnras] {10.1093/mnras/stx2569}, \href
  {http://adsabs.harvard.edu/abs/2018MNRAS.473.3525Z} {473, 3525}

\bibitem[\protect\citeauthoryear{{Zubovas}}{{Zubovas}}{2019}]{Zubovas2019MNRASa}
{Zubovas} K.,  2019, \mn@doi [\mnras] {10.1093/mnras/sty3211}, \href
  {https://ui.adsabs.harvard.edu/abs/2019MNRAS.483.1957Z} {483, 1957}

\bibitem[\protect\citeauthoryear{{Zubovas} \& {King}}{{Zubovas} \&
  {King}}{2012}]{Zubovas2012ApJ}
{Zubovas} K.,  {King} A.,  2012, \mn@doi [\apjl] {10.1088/2041-8205/745/2/L34},
  \href {http://adsabs.harvard.edu/abs/2012ApJ...745L..34Z} {745, L34}

\bibitem[\protect\citeauthoryear{{Zubovas} \& {King}}{{Zubovas} \&
  {King}}{2014}]{Zubovas2014MNRASa}
{Zubovas} K.,  {King} A.~R.,  2014, \mn@doi [\mnras] {10.1093/mnras/stt2472},
  \href {http://adsabs.harvard.edu/abs/2014MNRAS.439..400Z} {439, 400}

\bibitem[\protect\citeauthoryear{{Zubovas} \& {King}}{{Zubovas} \&
  {King}}{2016}]{Zubovas2016MNRASb}
{Zubovas} K.,  {King} A.,  2016, \mn@doi [\mnras] {10.1093/mnras/stw1845},
  \href {http://adsabs.harvard.edu/abs/2016MNRAS.462.4055Z} {462, 4055}

\bibitem[\protect\citeauthoryear{{Zubovas}, {Sabulis}  \& {Naujalis}}{{Zubovas}
  et~al.}{2014}]{Zubovas2014MNRASc}
{Zubovas} K.,  {Sabulis} K.,   {Naujalis} R.,  2014, \mn@doi [\mnras]
  {10.1093/mnras/stu1048}, \href
  {http://adsabs.harvard.edu/abs/2014MNRAS.442.2837Z} {442, 2837}

\bibitem[\protect\citeauthoryear{{Zubovas}, {Bourne}  \& {Nayakshin}}{{Zubovas}
  et~al.}{2016}]{Zubovas2016MNRASa}
{Zubovas} K.,  {Bourne} M.~A.,   {Nayakshin} S.,  2016, \mn@doi [\mnras]
  {10.1093/mnras/stv2971}, \href
  {http://adsabs.harvard.edu/abs/2016MNRAS.457..496Z} {457, 496}

\makeatother
\end{thebibliography}
\end{document}